\documentclass[preprintnumbers,notitlepage,a4paper,aps,prd,onecolumn,superscriptaddress,nofootinbib,groupedaddress]{revtex4}

\usepackage{amsmath}
\usepackage{amsfonts,color}
\usepackage{amsthm}
\usepackage{pdfpages}
\usepackage{empheq}
\usepackage{amssymb,float}
\usepackage{booktabs,multirow}
\usepackage{titlesec}
\usepackage{hyphenat}
\usepackage{ragged2e}
\usepackage{amsmath}
\usepackage{accents}
\usepackage[utf8]{inputenc}
\usepackage{color}
\usepackage{hyperref}
\usepackage{enumitem}
\usepackage{tikz}
\usepackage{comment}
\usepackage{ulem}
\usepackage{cancel}
\usepackage{graphicx}
\usepackage{mathrsfs}
\usetikzlibrary{shapes.geometric}
\usetikzlibrary{arrows.meta,arrows}

\titleformat{\paragraph}
{\normalfont\normalsize\bfseries}{\theparagraph}{1em}{}
\titlespacing*{\paragraph}
{0pt}{3.25ex plus 1ex minus .2ex}{1.5ex plus .2ex}

\allowdisplaybreaks
\hypersetup{
    colorlinks=true,
    linkcolor=blue,
    filecolor=magenta,      
    citecolor=blue
}

\allowdisplaybreaks[4] 
\begin{document}


\title{
Gauge-invariant cosmological perturbations in Type~3 New General Relativity and background-hierarchy bounds
}

\author{Kyosuke Tomonari}
\email[]{ktomonari.phys@gmail.com}
\affiliation{Institute of Astrophysics, Central China Normal University, Wuhan 430079, China}
\affiliation{Interfaculty Initiative in Information Studies, Graduate School of Interdisciplinary Information Studies, The University of Tokyo. 7-3-1 Hongo, Bunkyo-ku, Tokyo 113-0033, Japan}
\author{Daniel Blixt}
\email[]{blixtd@chalmers.se}
\affiliation{ Department of Mathematical Sciences, Chalmers University of Technology and University of Gothenburg, SE-412 96, Gothenburg, Sweden}
\author{Sebastian Bahamonde}
\email[]{sbahamondebeltran@gmail.com}
\affiliation{Cosmology, Gravity, and Astroparticle Physics Group, Center for Theoretical Physics of the Universe, Institute for Basic Science (IBS), Daejeon 34126, Korea}

\begin{abstract}
In this paper, we investigate background-hierarchy bounds in Type~3 of New General Relativity (NGR). These bounds arise when the contribution associated with the evolution of the background spacetime exceeds that of the kinetic term in the perturbed Lagrangian. 
Type~3 of NGR has two free parameters and is described in a pure-tetrad formulation while preserving diffeomorphism invariance and spatial rotations.
We first review Type~3 and identify preferable gauge choices for metric-affine gauge theories of gravity with Weitzenb\"ock connection, including NGR, from the viewpoint of symmetry in both Dirac--Bergmann analysis and linear perturbation theory. We then revisit the perturbative analysis of Type~3 and show that the propagating modes are correctly identified even when the perturbed Lagrangian is not written solely in terms of gauge-invariant variables. Finally, we derive the background-hierarchy bounds for the scalar, transverse-vector, and tensor modes around a flat FLRW background, and identify the region of parameter space in which the linear perturbation theory of Type~3 remains viable for cosmological applications. 
\end{abstract}

\maketitle

\section{\label{01} Introduction}

New General Relativity (NGR) is a three parameter extension of the Teleparallel Equivalent to General Relativity (TEGR), in which gravity is described by torsion rather than curvature in a parity-preserving manner~\cite{Einstein1928,Hayashi:1979qx}. 
The presence of additional degrees of freedom (DOFs) compared to TEGR motivates its study as a candidate framework to address outstanding cosmological puzzles, such as dark energy~\cite{SupernovaCosmologyProject:1998vns,SupernovaSearchTeam:1998fmf,Planck:2018nkj}, dark matter~\cite{Freese:2008cz,Billard:2021uyg,Planck:2018nkj}, and tensions in cosmological parameters~\cite{Planck:2018vyg,H0LiCOW:2019pvv,Schoneberg:2022ggi,Riess:2019cxk,ACT:2023kun}. 

To investigate these phenomenological issues, it is essential to clarify the nature of the degrees of freedom (DOFs) in NGR from the viewpoint of cosmological perturbations, based on its underlying constraint structure.
Recently, using the Dirac-Bergmann analysis (DB analysis)~\cite{Bergmann:1949zz,BergmannBrunings1949,Dirac:1950pj,Bergmann1950,Anderson:1951ta,Dirac:1958sq,Dirac:1958sc}, the authors revealed the full constraint structure and systematically counted the DOFs for all types of NGR~\cite{Blixt:2018znp,Tomonari:2024ybs,Tomonari:2024lpv}.
Furthermore, the authors revisited the framework of linear perturbation theory from the perspective of breaking local Lorentz symmetry in the context of DB analysis and performed a perturbative analysis around a flat FLRW background~\cite{Tomonari:2025kgb}. 
However, discrepancies remain regarding the counting of propagating modes~\cite{Golovnev:2023jnc} and the existence of ghost degrees of freedom~\cite{VanNieuwenhuizen:1973fi,Kuhfuss:1986rb,Golovnev:2023ddv,Mikura:2023ruz,Bahamonde:2024zkb}. 
It should be clarified that in this work the so-called Weitzenböck gauge (or pure-vierbein formulation) is assumed. 
The breaking of Lorentz symmetry (or Lorentz invariance) referred to in this work is the part that manifests itself in the DB analysis and corresponds to transformations of the vierbein (but not of the spin connection). 
In Ref.~\cite{Blixt:2022rpl}, this is defined as breaking of Lorentz invariance of Type~2, for instance. 
It is possible to reformulate the work covariantly by introducing the spin-connection \cite{Krssak:2015oua}. 
However, it would not affect the results of this work, since in NGR the effects of spin connection can be compensated with the method presented in Ref.~\cite{Blixt:2018znp}. 
Hence, we find it more convenient to work in the pure-vierbein formulation. 

For Types~1, 2, 3, 5, and 8 around the Minkowski background spacetime, two independent studies have investigated the propagating modes and instabilities~\cite{Bahamonde:2024zkb,Golovnev:2023ddv}.
The counting of propagating modes agrees between these works for all cases except for Type~8.
In Type~8, tensor modes are found in Ref.~\cite{Bahamonde:2024zkb}, whereas these modes are absent in Ref.~\cite{Golovnev:2023ddv}.
Regarding instabilities, these works also report different results. 
In Ref.~\cite{Bahamonde:2024zkb}, Types~1 and~5 suffer from instabilities, whereas all of these types are free from instabilities in Ref.~\cite{Golovnev:2023ddv}. 
For further discussions on instabilities, see Sec.~V of Ref.~\cite{Bahamonde:2024zkb} and references therein. 

For all Types~1–9 around a flat FLRW background spacetime, linear perturbative analyses have been performed in Refs.~\cite{Golovnev:2023jnc,Tomonari:2025kgb}. 
In Ref.~\cite{Tomonari:2025kgb}, we identified additional propagating modes for all types except for Types~1 and~6, which are absent in Ref.~\cite{Golovnev:2023jnc}. 
These discrepancies can be attributed to the following two points: 
whether or not 
\begin{enumerate}
    \item second-order terms of perturbation fields are included in the torsion and metric tensor,
    \item the gauge choice is performed in a manner consistent with the symmetry inherent in NGR, as clarified by the DB analysis.
\end{enumerate}
The results in Ref.~\cite{Tomonari:2025kgb} are summarized in Table~\ref{Table:PropertyOfNGR}. 
The first point has been already addressed in Ref.~\cite{Tomonari:2025kgb}. 
However, the second point requires further investigation, setting it for the first purpose of this paper.~\footnote{
In a FLRW background spacetime, the perturbed torsion tensor takes the form $T=[\mathrm{terms\,proportional\,to\,} H] + [\mathrm{1st\,order\,perturbation\,terms}] + [\mathrm{2nd\,order\,perturbation\,terms}] + \cdots$. 
Since $TT$ contains a term $[\mathrm{terms\,proportional\,to\,} H]\times[\mathrm{2nd\,order\,perturbation\,terms}]$, one must expand the perturbed torsion tensor up to second order. 
} 

In cosmological applications, Type~3 is of particular interest since Table~\ref{Table:PropertyOfNGR} indicates that the linear perturbation theory of Type~3 correctly captures all DOFs of the full theory. 
Type~6 is just an equivalent formulation to general relativity; there is no new content. 
Type~9 does not contain tensor modes so that this theory is not suitable to describe gravity. 
In contrast, perturbations of all other types, except for Types~3,~6 (TEGR), and 9, fail to describe the DOFs of the full theory, suggesting that these linearized models are not suitable for cosmological applications. 
As a note in comparison of Type~3 to Type~1, all recent works agree with the absence of ghost instability in Type~3 of NGR~\cite{Golovnev:2023jnc,Golovnev:2023ddv,Bahamonde:2024zkb,Tomonari:2025kgb}. 
On the other hand, although the linear perturbation of Type~1 NGR can capture all DOFs of the full theory correctly in a specific situation, some works indicate that this theory suffers from ghost instability~\cite{VanNieuwenhuizen:1973fi,Hayashi:1979qx,Deffayet:2023wdg}. 
Namely, there is no consensus on this perspective in Type~1 of NGR. Given this situation, we restrict our analysis to Type~3 of NGR.~\footnote{
We comment on Ref.~\cite{Golovnev:2026xjc}. 
In the strict vacuum Minkowski limit, the Type~3 vector sector does not possess a regular propagating vector mode. 
Although the equations admit a non-oscillatory solution, this solution is not suitable for cosmological applications. 
On an expanding FLRW background spacetime, the zeroth-order torsion proportional to $H$ generates additional terms in the quadratic action and in the equations of motion, which can activate background-induced cosmological vector modes. 
Therefore, the vector modes found in our FLRW analysis should be interpreted as background-induced modes whose Minkowski-limit structure degenerates. In detail, see Appendix~\ref{app:01}.
}

Type~3 of NGR is indeed a candidate for cosmological applications.
However, Refs.~\cite{Bahamonde:2024zkb} indicated that this type may suffer from strong coupling issues around a background spacetime.
As will be reviewed in Sec.~\ref{05-01} of the current paper, there are two distinct notions of strong coupling in theories of gravity.
In particle physics, since the background spacetime is fixed \textit{a priori} to be Minkowski, strong coupling typically refers to a situation in which the dominant contribution in the perturbative expansion switches from the kinetic term to higher-order interaction terms.
This is the conventional definition of strong coupling.
On the other hand, in gravitational theories, the background spacetime is not fixed in advance.
In cosmology, the background spacetime is dynamical and generally expanding.
This feature leads us to another possibility: 
the dominant contribution of a perturbation can switch from its kinetic term to background terms, which appear at zeroth order.
In this study, we investigate strong coupling in the latter case and call it a `\textit{perturbativity/background-hierarchy bound}' in order to distinguish it from the conventional Effective Field Theoretic (EFT) one.\footnote{The phenomenon of `perturbativity/background-hierarchy bound' is well-known  \cite{JimenezCano:2021rlu}, however, it is generally just referred to as `strong coupling'.} 
Cosmological perturbation theory is viable only in the region in which propagating modes do not suffer from the background-hierarchy bound. 
The second purpose of this paper is to investigate the viability of the perturbed models of Type~3 NGR. 

This paper is organized as follows.
In Sec.~\ref{02}, we review Type~3 of NGR with emphasis on its symmetry structure.
In Sec.~\ref{03}, we clarify the relation between the symmetries revealed by the DB analysis and those appearing in linear perturbation theory.
Based on this relation, we introduce preferable gauge choices to analyse metric-affine gauge theories of gravity (MAG) with Weitzenb\"{o}ck gauge, including NGR as a special case. 
Our results are summarized in Table~\ref{Table:PropagatingModesImPreferableGaugeChoice}.
In Sec.~\ref{04}, we revisit the previous analysis given in Ref.~\cite{Tomonari:2025kgb} by performing it in an \textit{un}preferable gauge and explicitly demonstrate how an incorrect conclusion arises. 
In Sec.~\ref{05}, after reviewing the concept of strong coupling in particle physics using the Proca theory as a representative example to clarify the difference of background-hierarchy bound from the conventional EFT strong coupling, we investigate the background-hierarchy bound of Type~3 of NGR around a flat FLRW background.
Moreover, we identify the condition under which the theory can remain predictive beyond the background-hierarchy bound, with the aim of applying it to cosmology. 
In Sec.~\ref{06}, we conclude this work and indicate future direction. 

In this paper, we use the unit $c^{4}/16\pi G=1$, where $c$ and $G$ are the speed of light and Newton's constant, respectively.
In Sec.~\ref{05}, we use the reduced Planck mass, denoted by $M_{pl}$, except for Sec.~\ref{05-01}.
In Sec.~\ref{05-01}, we use the units $c=1$ and $\hbar=1$, where $\hbar$ is the reduced Planck constant.
Greek indices $\alpha,\beta,\gamma,\cdots,\mu,\nu,\rho,\cdots$ denote spacetime indices.
Capital Latin indices $A,B,C,\cdots,I,J,K,\cdots$ denote internal-space indices.
Lowercase Latin indices $a,b,c,\cdots$ and $i,j,k,\cdots$ denote internal-space spatial indices and spacetime spatial indices, respectively.
All calculations in this work were performed using Cadabra~\cite{Peeters:2007wn}, a free and powerful computer algebra system.

\begin{table}[ht!]
    \centering
    \renewcommand{\arraystretch}{1.5}
    \begin{tabular}{ c || c | c | c | c | c }
	Theory 
    &
    \begin{tabular}{c}
    Parameter-space \\
    conditions
    \end{tabular}
    & 
    Regularity 
    &
    \begin{tabular}{c}
        \# of Non-linear DOF \\ 
        (DB analysis)
    \end{tabular} 
    & 
    \begin{tabular}{c}
        Propagating Modes \\ 
        (Perturbative analysis \\
        around flat FLRW spacetime) 
    \end{tabular}  
    & 
    \begin{tabular}{c}
    Ghost-free \\
    conditions
    \end{tabular}
    \\ \hline \hline
	Type~1 & arbitrary & $-$ & 8 & \begin{tabular}{c}6 - 8: \\($h_{ij}, \alpha, \tilde{\sigma}, \alpha_{i} ($or $\tilde{V}_{i})$; $\tilde{V}_{i}($or $\alpha_{i})$) \end{tabular}& \begin{tabular}{c} $2 c_{1} - c_{2} < 0$ \& \\ $2 c_{1} - c_{2} + c_{3} > 0$ \& \\ $2 c_{1} + c_{2} < 0$ \end{tabular} \\ \hline
	Type~2 & $2c_{1} - c_{2} + c_{3} = 0$ & $\checkmark$ & 6 & 5: ($h_{ij}, \tilde{\sigma}, \tilde{V}_{i}$) & \begin{tabular}{c} $2 c_{1} - c_{2} < 0$ \& \\ $2 c_{1} + c_{2} <0$ \end{tabular} \\ \hline
	Type~3 & $2 c_{1} + c_{2} = 0$ & $\checkmark$ & 5 & 5: ($h_{ij}, \alpha, \alpha_{i}$) & \begin{tabular}{c} $2 c_{1} - c_{2} < 0$ \& \\ $2 c_{1} - c_{2} + c_{3} > 0$	\end{tabular}\\ \hline
	Type~4 & $2 c_{1} - c_{2} = 0$ & $\times$ & 5 & 2 - 4: ($\alpha, \tilde{\sigma}$; either $\alpha_{i}$ or $\tilde{V}_{i}$) & \begin{tabular}{c} $2 c_{1} - c_{2} + c_{3} > 0$ \& \\ $2 c_{1} + c_{2} < 0$ \end{tabular}\\ \hline
	Type~5 & $2 c_{1} - c_{2} + 3 c_{3} = 0$ & $\checkmark$ & 7 & 6: ($h_{ij}, \alpha, \tilde{\sigma},$ either $\alpha_{i}$ or $\tilde{V}_{i}$) & \begin{tabular}{c} $2 c_{1} - c_{2} < 0$ \& \\ $2 c_{1} - c_{2} + c_{3} > 0$ \& \\ $2 c_{1} + c_{2} < 0$ \end{tabular}\\ \hline
	Type~6 & \begin{tabular}{c} $2\,c_{1} - c_{2} + c_{3} = 0$ \& \\ $2\,c_{1} + c_{2} = 0$\end{tabular} & $\checkmark$ & 2 & 2: ($h_{ij}$) & $2 c_{1} - c_{2} < 0$ \\ \hline 
	Type~7 & \begin{tabular}{c} $2\,c_{1} + c_{2} = 0$ \& \\ $2\,c_{1} - c_{2} = 0$\end{tabular} & $\times$ & 0 (Topological) & 3: ($\alpha, \alpha_{i}$) & \begin{tabular}{c} $2 c_{1} - c_{2} + c_{3} > 0$ \end{tabular}\\ \hline
	Type~8 & \begin{tabular}{c} $2\,c_{1} + c_{2} = 0$ \& \\ $2\,c_{1} - c_{2} + 3\,c_{3} = 0$\end{tabular} & $\checkmark$ & 6 or 4 (Bifurcate) & 5: ($h_{ij}, \alpha, \alpha_{i}$) & \begin{tabular}{c} $2 c_{1} - c_{2} < 0$ \& \\ $2 c_{1} - c_{2} + c_{3} > 0$	\end{tabular}\\ \hline
	Type~9 & \begin{tabular}{c}$2\,c_{1} - c_{2} + c_{3} = 0$ \& \\ $2\,c_{1} - c_{2} = 0$ \& \\ $2\,c_{1} - c_{2} + 3\,c_{3} = 0$ 
    \end{tabular} & $\checkmark$ & 3 & 3: ($\tilde{\sigma}, \tilde{V}_{i}$) & $2 c_{1} + c_{2} < 0$ \\ \hline
    \end{tabular}
    \caption{Degrees of freedom in each Type of New General Relativity. }
    \label{Table:PropertyOfNGR}
\end{table}

\section{\label{02} Type~3 of New General Relativity}

New General Relativity (NGR) is an extension of the Teleparallel Equivalent of General Relativity (TEGR) with three free parameters. 
TEGR itself is a specific class of metric-affine gauge theories of gravity (MAG) and is equivalent to General Relativity (GR) up to a boundary term in the action. 
One can study these theories in Refs.~\cite{Hehl:1994ue,Bahamonde:2021gfp}.

In TEGR and NGR, the affine connection is expressed in terms of the vierbein and co-vierbein fields, $e_{I}{}^{\mu}$ and $e^{I}{}{\mu}$, as\begin{equation}
    \Gamma{}^{\rho}_{\mu\nu} 
    = 
    e_{I}{}^{\rho}\,\partial_{\mu}\,e^{I}{}_{\nu}
    \,,
\label{Weitzenboeck connection}
\end{equation}
and thus, the torsion tensor, $T^{\rho}{}_{\mu \nu}$, is represented by 
\begin{equation}
    T^{\rho}{}_{\mu\nu} 
    = 
    2 \Gamma^{\rho}_{[\mu\nu]} 
    = 
    e_{I}{}^{\rho}\,\left(\,\partial_{\mu}e^{I}{}_{\nu} 
    - \partial_{\nu}e^{I}{}_{\mu}\,\right)
    \,.
\label{torsion in W-gauge}
\end{equation}
Here, the general form of an affine connection is given by 
\begin{equation}
    \Gamma^{\rho}_{\mu\nu} 
    = 
    e_{I}{}^{\rho}\,\partial_{\mu}\,e^{I}{}_{\nu} + \omega^{I}_{J\mu} e^{J}{}_{\nu} e_{I}{}^{\rho} \,.
\label{affine connection}
\end{equation}
The Weitzenb\"ock gauge is $\omega^{I}_{J\mu} = 0$. 
Inheriting this property, we call Eq.~\eqref{Weitzenboeck connection} the Weitzenb\"{o}ck connection. 

The Lagrangian of NGR consists of the torsion tensor $T^{\rho}{}_{\mu \nu}$ as follows~\cite{Hayashi:1979qx}:
\begin{equation}
\begin{split}
    L_\mathrm{NGR}
    := 
    \theta^{-1}\,\mathcal{L}_\mathrm{NGR} 
    &= c_{1}\,T^{\mu\nu\rho}\,T_{\mu\nu\rho} + c_{2}\,T^{\mu\nu\rho}\,T_{\rho\mu\nu} + c_{3}\,T^{\mu}{}_{\mu\rho}\,T_{\nu}{}^{\nu\rho} \,\\
    &= c_{1}\,g_{\mu\sigma}\,g^{\nu\lambda}\,g^{\rho\kappa}\,T^{\mu}{}_{\lambda\kappa}\,T^{\sigma}{}_{\nu\rho} + c_{2}\,g^{\nu\lambda}\,T^{\mu}{}_{\lambda\rho}\,T^{\rho}{}_{\mu\nu} + c_{3}\,g^{\rho\nu}\,T^{\mu}{}_{\mu\rho}\,T^{\lambda}{}_{\lambda\nu} \,,
\end{split}
\label{Lagrangian of NGR}
\end{equation}
where $\theta$ denotes the determinant of the co-vierbein field components, and $c_{1}$, $c_{2}$, and $c_{3}$ are three real-valued free parameters. 
For example, in TEGR, which is equivalent to GR up to a boundary term in the action integral (although the underlying spacetime geometry differs), the parameters take the values $c_{1} = -1/4$, $c_{2} = 1/2$, and $c_{3} = 1$. 

Varying the NGR Lagrangian with respect to the vierbein, we obtain the field equations as follows~\cite{Golovnev:2023uqb,Bahamonde:2024zkb,Tomonari:2025kgb}\footnote{
Several groups have derived the field equations. 
All results coincides with one another, but the form is different. 
We use the form that is provided in our previous work~\cite{Tomonari:2025kgb}. 
}:
\begin{equation}
\begin{split} 
    \frac{1}{2}\,e_{A}{}^{\rho}\,\mathcal{T}_{\rho}{}^{\nu} 
    &=
    \theta^{-1}\,\partial_{\mu}(\theta\,e_{A}{}^{\rho}\,S_{\rho}{}^{\mu\nu}) 
    + e_{A}{}^{\lambda}\,T^{\rho}{}_{\mu\lambda}\,S_{\rho}{}^{\mu\nu} 
    - \frac{1}{4}\,e_{A}{}^{\nu}\,L_\mathrm{NGR} 
    \,,\\
    S_{\rho}{}^{\mu\nu} 
    &:= 
    c_{1}\,T_{\rho}{}^{\mu\nu} 
    + c_{2}\,T^{[\mu\nu]}{}_{\rho} 
    + c_{3}\,\delta_{\rho}{}^{[\mu}T^{|\kappa|}{}_{\kappa}{}^{\nu]} 
    \,.
\end{split}
\label{field equation of NGR}
\end{equation}
Here, $e_{A}{}^{\rho}\,\mathcal{T}_{\rho}{}^{\nu} = \theta^{-1}\,\delta\mathcal{L}_\mathrm{matter}\,/\,\delta\,e^{A}{}_{\nu}$ denotes the energy-momentum tensor pushed forward by the vierbein field.

NGR models are classified by the parameter space $(c_{1}, c_{2}, c_{3})$. 
For Type~3, we impose the condition~\cite{Blixt:2018znp}
\begin{equation}
    2 c_{1} + c_{2} = 0 
\label{Type 3 parameter condition}
\end{equation}
on the parameter space. 

In general, NGR admits a Hamiltonian formulation and can be decomposed into nine independent types according to the combination of primary constraints which are related to the $SO(3)$-irreducible representations of the canonical momenta~\cite{Blagojevic:2000qs,Blixt:2018znp}\footnote{A rigid $SO(1,3)$-irreducible decomposition of the \textit{torsion} tensor is presented in Ref.~\cite{Hayashi:1979qx}.}.
The total Hamiltonian of Type~3 is given by 
\begin{equation}
\begin{split}
    &\mathcal{H} 
    = 
    \mathcal{H}_{0} 
    + \sqrt{h}\ {}^\mathcal{A}\lambda_{ij} {}^{\mathcal{A}}C^{ij} + D_i[\pi_A{}^i(\alpha \xi^{A}+\beta^{j}e^A{}_j )]  
    \,,\\
    &\mathcal{H}_{0}
    =
    {}^\mathcal{V}\mathcal{H} + {}^\mathcal{S}\mathcal{H} + {}^\mathcal{T}\mathcal{H} \\
    &\qquad - \alpha \Big(\sqrt{h}\ {}^{3}\mathbb{T} - \xi^A D_i\pi_A{}^i \Big) - \beta^k\Big(T^A{}_{jk}\pi_A{}^j + e^A{}_k D_{i}\pi_A{}^i\Big) + {}^{\alpha}\lambda\pi_{0} + {}^{\beta}\lambda^{i}\pi_{i} 
    \,.
\end{split}
\label{total Hamiltonian of type 3}
\end{equation}
where $D_{i}$ denotes the covariant derivative on each leaf of the ADM foliation~\cite{Deser:1959vvc,Arnowitt:1960es,Arnowitt:1962hi}, and ${}^{\mathcal{A}}\lambda_{ij}$, ${}^{\alpha}\lambda$, and ${}^{\beta}\lambda$ are the Lagrange multipliers. 
$\alpha$ and $\beta_{i}$ are the lapse function and the shift vector in ADM foliation, respectively.
The canonical momenta conjugate to $\alpha$ and $\beta_{i}$ are denoted by $\pi_{0}$ and $\pi_{i}$, respectively.
$\pi_A{}^i$ is the conjugate momentum to $e^{A}{}_{j}$. 
The quantities $\xi_{A}$ and $h_{ij}$ are the unit normal vector to, and the induced spatial metric on, each leaf of the ADM foliation, respectively.
The explicit expressions of ${}^{3}\mathbb{T}$, $\mathcal{H}_{0}$, ${}^{\mathcal{A}}C^{ij}$, ${}^\mathcal{V}\mathcal{H}$, ${}^\mathcal{S}\mathcal{H}$, and ${}^\mathcal{T}\mathcal{H}$ are found in the original work~\cite{Blixt:2018znp}. 

Type~3 possesses the constraint
\begin{equation}
    {}^{\mathcal{A}}C^{ij} \approx 0 
    \,.
\end{equation}
The Poisson bracket algebra (PB algebra) of ${}^{\mathcal{A}}C^{ij}$ is calculated as~\cite{Tomonari:2024ybs}
\begin{equation}
    \{{}^\mathcal{A}\mathcal{C}^{ij}(t\,,\vec{x})\,,{}^\mathcal{A}\mathcal{C}^{kl}(t\,,\vec{y})\}_{P.b.} 
    = 
    \frac{2}{\sqrt{h}}\left(\delta^{j[l}{}^{\mathcal{A}}\mathcal{C}^{k]i} + \delta^{i[k}{}^{\mathcal{A}}\mathcal{C}^{l]j}\right)\delta^{(3)}(\vec{x}-\vec{y}) 
    \,,
\label{PB-algebra of AC and AC}
\end{equation}
where $\{\cdot,\cdot\}_{P.b.}$ denotes the Poisson bracket. 
This algebra corresponds to $so(3)$ Lie algebra. 

For diffeomorphism symmetry, the gauge generator is constructed as follows~\cite{Tomonari:2024ybs}:
\begin{equation}
    \mathcal{G}_{\rm Diffeo} = g_{(1)}^{\mu}\phi^{(1)}_{\mu}+g_{(2)}^{\mu}\phi^{(2)}_{\mu}\,,
\label{Gauge genrator for Diffeomorphism}
\end{equation}
where $\phi^{(1)}_{\mu}:=(\pi_{0}\,,\pi_{i})$ and $\phi^{(2)}_{\mu}=(\phi^{(2)}_{0}\,,\phi^{(2)}_{i})$ are the primary and secondary constraints of Type~3, respectively, which satisfy the hypersurface deformation algebra (HDA algebra)~\cite{Dirac:1958sc}. 
(See also Sec.~\ref{03-01}.)
Here, $g^{(1)}_{\mu}$ and $g^{(2)}_{\mu}$ are arbitrary functions. 
For the constraint ${}^{\mathcal{A}}C^{ij} \approx 0$, the corresponding gauge generator is given by~\cite{Tomonari:2024ybs}
\begin{equation}
    \mathcal{G}_{\rm Type\,3} = g^{i}L_{i}\,,\quad L_{i}=\frac{1}{2}\epsilon_{ijk}\,{}^{\mathcal{A}}\mathcal{C}^{jk}\,,
\label{Gauge generator for Type 3}
\end{equation}
where $g^{i}$ are arbitrary functions. 

To examine the gauge symmetries, we verify the invariance of the total Hamiltonian under the gauge transformations generated by Eqs.~\eqref{Gauge genrator for Diffeomorphism} and~\eqref{Gauge generator for Type 3}, respectively. 
The gauge transformations of the total Hamiltonian, Eq.~\eqref{total Hamiltonian of type 3}, with respect to each generator, Eqs.~\eqref{Gauge genrator for Diffeomorphism} and~\eqref{Gauge generator for Type 3}, are respectively given by~\cite{Tomonari:2024ybs}
\begin{equation}
    \delta_{\rm Diffeo}\tilde{\mathcal{H}}_{\rm Type\,3} = \left\{\tilde{\mathcal{H}}_{\rm Type\,3}\,,\mathcal{G}_{\rm Diffeo}\right\}_{P.b.} = -\mathcal{L}_{\mathcal{G}_{\rm Diffeo}}\tilde{\mathcal{H}}_{\rm Type\,3}\approx0\,
\label{gauge transformation wrt diffeo}
\end{equation}
and
\begin{equation}
    \delta_{\rm Type\,3}\tilde{\mathcal{H}}_{\rm Type\,3} = \left\{\tilde{\mathcal{H}}_{\rm Type\,3}\,,\mathcal{G}_{\rm Type\,3}\right\}_{P.b.} = -\mathcal{L}_{\mathcal{G}_{\rm Type\,3}}\tilde{\mathcal{H}}_{\rm Type\,3}\approx0\,
\label{gauge transformation wrt rotation}
\end{equation}
as desired, where $\mathcal{L}_{X}$ denotes the Lie derivative with respect to the vector field $X$. 
These results indicate the consistency of the theory; 
that is, Type~3 NGR is invariant under diffeomorphism and spatial rotation symmetries.

\section{\label{03}Gauge symmetry in theory of gravity versus its linear perturbation theory}

\subsection{\label{03-01} Gauge symmetry in theory of gravity and degrees of freedom} 

DB analysis reveals not only the propagating DOFs but also the symmetries of a theory. 
For instance, the DB analysis ensures that GR possesses two symmetries: diffeomorphism invariance and local Lorentz symmetry Refs~\cite{Henneaux:1983vi,Flinckman:2026kpw}.
As shown in Refs.~\cite{Tomonari:2024ybs,Tomonari:2024lpv}, in NGR the diffeomorphism symmetry always holds, whereas the local Lorentz symmetry is broken in all cases except for Type~6 (TEGR);
only Type~6 (TEGR) is equivalent to GR with respect to both symmetries.

Symmetry breaking can give rise to new DOFs.
This statement can be understood from the definition of DOFs as follows~\cite{Shanmugadhasan1973,MaskawaNakajima1976,Dominici:1979bg,Tomonari:2023vgg}:
\begin{equation}
    \mathrm{DOFs}  = \frac{1}{2} \big[ 
    \mathrm{\{\#\,of\,phase\,space\,dimension}\} - 2 \times \mathrm{\{\#\,of\,first\text{-}class\,constraint(s)\}} - \mathrm{\{\#\,of\,second\text{-}class\,constraints\}}
    \big],
\label{Def of DOFs}
\end{equation}
where the symbol ``$\#$'' denotes the number. 
As a remark on the construction of constraints, the first-class constraints must be chosen as a functionally independent maximal set.
Then, the number of second-class constraints turns out to be always even~\cite{Sugano:1982bm}. 
The crucial point here is that a set of first-class constraints generates the gauge symmetries of a theory, and each symmetry is characterized by a proper PB algebra~\cite{Sugano:1982bm, Sugano:1986xb, Sugano:1989rq, Sugano:1991ke, Sugano:1991kd}.
This property implies that when a gauge symmetry is broken, the number of first-class constraints changes, which in turn suggests that the number of DOFs can increase.
Such symmetry breaking typically occurs when an original theory is modified or extended.
For example, NGR, as an extension of the original theory, TEGR, no longer respects the local Lorentz symmetry in PB-algebra~\cite{Tomonari:2024ybs,Tomonari:2024lpv}.

There are three typical patterns by which the number of DOFs increases, as summarized below:
\begin{enumerate}
    \item Pattern   I. A first-class constraint disappears, giving rise to one new DOF; 
    \item Pattern  II. Two first-class constraints become second-class, giving rise to one new DOF; 
\end{enumerate}
Patterns I and II occur in modifications or extensions of an original theory, and these two patterns are directly related to the emergence of new DOFs in NGR.

We mention two important representations of symmetry in MAG theories. 
The first is diffeomorphism symmetry, which is represented by the HDA algebra as follows~\cite{Dirac:1958sc}:
\begin{equation}
\begin{split}
    &\{\phi^{(2)}_{i}(t\,,\vec{x})\,,\phi^{(2)}_{j}(t\,,\vec{y})\}_{P.b.} = \left(\phi^{(2)}_{j}(t\,,\vec{x})\partial^{(x)}_{i}-\phi^{(2)}_{i}(t\,,\vec{y})\partial^{(y)}_{j}\right)\delta^{(3)}(\vec{x}-\vec{y})\,,\\
    \quad
    &\{\phi^{(2)}_{i}(t\,,\vec{x})\,,\phi^{(2)}_{0}(t\,,\vec{y})\}_{P.b.} = \phi^{(2)}_{0}(t\,,\vec{x})\partial^{(x)}_{i}\delta^{(3)}(\vec{x}-\vec{y})\,,\\
    &\{\phi^{(2)}_{0}(t\,,\vec{x})\,,\phi^{(2)}_{0}(t\,,\vec{y})\}_{P.b.} = \left(h^{ij}(t\,,\vec{x})\phi^{(2)}_{j}(t\,,\vec{x})+h^{ij}(t\,,\vec{y})\phi^{(2)}_{j}(t\,,\vec{y})\right)\partial^{(x)}_{i}\delta^{(3)}(\vec{x}-\vec{y})\,.
\end{split}
\label{HDA}
\end{equation}
The constraints $\phi^{(2)}_{\mu}\approx0$ $(\mu=1,2,3,4)$ are classified as secondary first-class constraints.\footnote{
Corresponding to each $\phi^{(2)}_{\mu}$, there exists a primary first-class constraint $\phi^{(1)}_{\mu}\approx0$.
The consistency condition for each $\phi^{(1)}_{\mu}$ gives rise to $\phi^{(2)}_{\mu}$.
}
The explicit forms of these constraints depend on the specific theory under consideration, whereas the PB algebra itself is universal among all diffeomorphism-invariant theories.
Therefore, if this symmetry is violated according to Pattern I shown above, up to eight DOFs can emerge. 
In the case of Pattern II, up to four DOFs can emerge.

In the case of TEGR, which is the original theory underlying NGR, this PB algebra is satisfied~\cite{Ferraro:2016wht,Ferraro:2018tpu}.
NGR is also known to preserve this symmetry~\cite{Tomonari:2024ybs,Tomonari:2024lpv}.

The second symmetry is the internal space symmetry, which is represented by a Lie algebra $\mathfrak{g}$ given as
\begin{equation}
\begin{split}
    &
    \{E_{I}(t\,,\vec{x})\,,E_{J}(t\,,\vec{y})\}_{P.b.} = f_{IJK}\,E_{K}(t\,,\vec{x})\,\delta^{(3)}(\vec{x} - \vec{y}) \,,
\end{split}
\label{symmetry G}
\end{equation}
where $E_{I}$ is the generator of this algebra and $f_{IJK}$ denotes its structure constants.
A semi-direct product of $\mathfrak{g}$ with a commutative algebra $\mathfrak{h}$, denoted by $\mathfrak{h} \rtimes \mathfrak{g}$, forms a generic internal-space symmetry of which PB-algebra is given by
\begin{equation}
\begin{split}
    &
    \{E_{I}(t\,,\vec{x})\,,P_{J}(t\,,\vec{y})\}_{P.b.} = g_{IJK}\,P_{K}(t\,,\vec{x})\,\delta^{(3)}(\vec{x} - \vec{y}) \,
    \\
    &
    \{P_{I}(t\,,\vec{x})\,,P_{J}(t\,,\vec{y})\}_{P.b.} = 0 
    \,,
\end{split}
\label{symmetry T rtimes G}
\end{equation}
in addition to Eq.~\eqref{symmetry G}, where $P_{I}$ is the generator of $\mathfrak{h}$ and $g_{IJK}$ are the structure constants of $\mathfrak{h} \rtimes \mathfrak{g}$.
The explicit representations of the constraints $E_{I}$ and $P_{I}$ depend on the specific theory under consideration, whereas the algebra itself is universal among all internal-space-invariant theories.

As a special case, this algebra reproduces the local Lorentz algebra $so(1,3)$, which contains $so(3)$ Lie algebra as a sub-algebra that appears in Type~3 (See Eq.~\eqref{PB-algebra of AC and AC}):
\begin{equation}
    \{\mathcal{C}^{(1)}_{IJ}(t\,,\vec{x})\,,\mathcal{C}^{(1)}_{KL}(t\,,\vec{y})\}_{P.b.} 
    = \Big({\eta_{JL}}\,\mathcal{C}^{(1)}_{IK}(t\,,\vec{x}) + {\eta_{IK}}\,\mathcal{C}^{(1)}_{JL}(t\,,\vec{x}) - {\eta_{JK}}\,\mathcal{C}^{(1)}_{IL}(t\,,\vec{x}) - {\eta_{IL}}\,\mathcal{C}^{(1)}_{JK}(t\,,\vec{x}) \Big)\delta^{(3)}(\vec{x} - \vec{y})\,. 
\label{local LI}
\end{equation}
In this case, $E_{I}$ corresponds to $\mathcal{C}^{(1)}_{IJ}$. 
The constraints $\mathcal{C}^{(1)}_{IJ} \approx 0$ $(I, J = 0,1,2,3)$ are classified as primary first-class constraints.
If this symmetry is broken according to Pattern I, up to six DOFs can emerge.
In the case of Pattern II, up to three DOFs can emerge.

TEGR satisfies this PB algebra~\cite{Ferraro:2016wht,Ferraro:2018tpu}, whereas in NGR this symmetry is violated in general, except for Type~6.
In particular, Type~3 of NGR, which plays a central role in the present investigation, preserves $so(3)$ symmetry.
This implies that in Type~3 the pseudo-scalar and pseudo-vector modes \textit{cannot} propagate, as shown in previous work~\cite{Tomonari:2025kgb}. 

\subsection{\label{03-02} Gauge symmetry in linear perturbation theory and degrees of freedom} 

The vierbein field components $e^{I}{}_{\mu}$ can be linearly expanded as~\cite{Dent:2010nbw,Chen:2010va,Wu:2012hs,Izumi:2012qj,Tomonari:2025kgb}\footnote{
Strictly speaking, these components are ``co-''vierbein field components. 
However, in much of the literature, they are commonly referred to as \textit{vierbein} or \textit{tetrad} field components. 
In the present paper, we follow this conventional terminology and simply refer to $e^{I}{}_{\mu}$ as the \textit{vierbein} field components. 
}
\begin{equation}
\begin{split}
    &e^{0}{}_{\mu} 
    = 
    \left(1 + \psi\right)\,\delta^{0}{}_{\mu} + a\,\left(\,\partial_{i}\alpha + \alpha_{i}\,\right)\,\delta^{i}{}_{\mu} 
    \,,\\
    &e^{a}{}_{\mu} 
    = 
    a\,\left(1 - \varphi\right)\,\delta^{a}{}_{\mu} 
    + \delta^{ai}\,(\partial_{i}F + G_{i})\,\delta^{0}{}_{\mu} 
    \\
    & \qquad \qquad
    + a\,\delta^{ai}\,
    \left[
        h_{ji} + \partial_{j}\partial_{i}B + C_{i j} + \epsilon_{ijk}\,\delta^{kl}\,\left(\partial_{l}\tilde{\sigma} + \tilde{V}_{l}\right)
    \right]
    \,\delta^{j}{}_{\mu} 
    \,,
\end{split} 
\label{vierbein linear perturbations}
\end{equation}
where we set $C_{i j} := 2 \partial_{(i}C_{j)}$.\footnote{
A resemble formulation to Refs.~\cite{Dent:2010nbw,Chen:2010va,Wu:2012hs,Izumi:2012qj} has been independently established in Ref.~\cite{Golovnev:2018wbh}. 
The correspondence between ours revisited in Ref.~\cite{Tomonari:2025kgb} and them can be listed as follows: 
\begin{equation}
\begin{split}
    &
    \varphi \leftrightarrow \psi\,, 
    \quad 
    \psi \leftrightarrow \phi\,, 
    \quad 
    B \leftrightarrow \sigma\,, 
    \quad 
    F \leftrightarrow \zeta\,, 
    \quad 
    \alpha \leftrightarrow \beta \,, 
    \quad 
    \tilde{\sigma} \leftrightarrow s \,, \\
    &
    C_{i} \leftrightarrow c_{i} \,,
    \quad
    G_{i} \leftrightarrow v_{i} \,, 
    \quad 
    \alpha_{i} \leftrightarrow u_{i} \,, 
    \quad 
    \tilde{V}_{i} \leftrightarrow \chi_{i} \,.
\end{split}
\end{equation}
The tensor mode is commonly introduced in both formulations. 
In our formulation, $C_{i}$ is introduced as $C_{ij} := 2 \partial_{(i}{C_{j)}}$ according to the convention, by contrast, they introduce it as $c_{i}$. 
In diffeomorphims invariant theories, $G_{i}$ does always not propagate; 
that is, we can always identify $\alpha_{i}$ to both $\mathcal{M}_{i} = ( u_{i} - v_{i} ) / 2$ and $\mathcal{L}_{i} = ( u_{i} + v_{i} ) / 2\,$. 
}
This perturbative decomposition can describe, at most, $10$ DOFs from the spacetime sector ($h_{ij}$, $\psi$, $\varphi$, $F$, $B$, $G_{i}$, $C_{i}$) and $6$ DOFs from the internal-space sector ($\alpha$, $\alpha_{i}$, $\tilde{\sigma}$, $\tilde{V}_{i}$), amounting to 16 candidate propagating modes.  
This framework can be applied to MAG theories with the Weitzenb\"{o}ck gauge formulated solely in terms of a vierbein field. 
NGR is one such theory.

For an infinitesimal coordinate transformation $x^{\mu} \rightarrow x'^{\mu} = x^{\mu} + \xi^{\mu}(x)\,$, the variation of the vierbein is given by
\begin{equation}
\begin{split}
    & 
    e^{I}{}_{\mu} \rightarrow e'^{I}{}_{\mu} 
    = e^{I}{}_{\mu} + \delta_{\xi}e^{I}{}_{\mu} 
    = e^{I}{}_{\mu} -\xi^{\nu}\partial_{\nu}e^{I}{}_{\mu} -e^{I}{}_{\nu}\,\partial_{\mu}\xi^{\nu} 
    \,,
\end{split}
\label{gauge transformation in linear perturbation}
\end{equation}
where $\delta_{\xi}\,e^{I}{}_{\mu}$ denotes the Lie derivative of $e^{I}{}_{\mu}$ with respect to $\xi^{\mu}$. 
Assuming that Eq.~\eqref{vierbein linear perturbations} is invariant under this transformation, we can derive the gauge transformation laws of the perturbation fields.

In practical applications, fixing a gauge greatly simplifies perturbative analysis. 
Since the gauge transformation is generated by four functions $\xi^{\mu}$, which can be decomposed into a temporal component $\xi^{0}$, a scalar component $\xi$, and a transverse vector component $\xi^{(v)}_{i}$, we can fix at most four perturbation fields. 
That is, our perturbation theory can describe a theory of gravity with Weitzenb\"{o}ck gauge using at most twelve independent perturbation fields. 
However, gauge fixing entails a subtle issue: 
its consistency with the symmetries revealed by the DB analysis of the underlying gravitational theory. 
In principle, one should not impose a gauge condition that eliminates or constrains a perturbation field associated with a symmetry identified by DB analysis. 
Otherwise, gauge fixing may inadvertently modify the original theory already at the linear level, and one ends up analyzing a different model with additional or spurious modes, which is merely motivated by the original theory rather than equivalent to it. 
Before counting propagating modes, we first discuss a preferable strategy for gauge fixing from a general viewpoint. 

First, we establish a criterion for classifying gauge transformations. 
To this end, we consider the gauge transformations of $\varphi$ and $\psi$~\cite{Izumi:2012qj,Tomonari:2025kgb}:
\begin{equation}
\begin{split}
    &\varphi' = \varphi + H \xi^{0} \,,\\
    &\psi' = \psi - \dot{\xi}^{0} \,.
\end{split}
\label{gauge transformation varphi and psi}
\end{equation}
We refer to this pair of perturbation fields as the $(\varphi,\psi)$ sector. 
By fixing the gauge parameter $\xi^{0}$, we can make either $\varphi$ or $\psi$ vanish. 

From the first transformation law, we can impose $\varphi' = 0$ by choosing 
\begin{equation}
    \xi^{0} = - H^{-1} \varphi \,.
\end{equation} 
With this gauge choice, the transformation of $\psi$ becomes 
\begin{equation}
    \psi' = \psi + \dot{H} H^{-2} \varphi - H^{-1} \dot{\varphi} \,;
\end{equation}
in this case, $\psi'$ is expressed in a well-defined manner solely in terms of the variables of the original theory. 
Alternatively, one may attempt to fix the gauge by imposing $\psi' = 0$, which formally requires
\begin{equation}
    \xi^{0} \overset{?}{=} \int \psi(t, \vec{x}) dt + \Xi(\vec{x}) \,,
\label{integral of xi0}
\end{equation}
where $\Xi(\vec{x})$ is an arbitrary function that depends only on spatial coordinates. 
However, the existence of such a function $\xi^{0}$ is not always guaranteed in general. 
If $\xi^{0}$ does not exist, the transformed variable $\varphi'$ cannot be represented as a well-defined function of the variables in the original theory, and the gauge transformation fails. 
This gauge choice is therefore not preferable, as it leads to an ill-defined gauge fixing. 
In particular, for numerical analysis, it is better to choose the gauge locally in time. 
Consequently, it is generally preferable to choose the gauge $\xi^{0}$ such that $\varphi' = 0$. 
From this observation, we adopt the following criterion for selecting a preferable gauge:
\begin{quote}
    \textit{A preferable gauge is one that is determined by solving, `purely algebraically', a condition that makes a perturbation field vanish.}
\end{quote}
In the present example, the gauge $\psi' = 0$ is not preferable, since it requires solving a partial differential equation,
$\psi(t,\vec{x}) - \dot{\xi}^{0}(t,\vec{x}) = 0$, 
whereas the gauge $\varphi' = 0$ is preferable, as it only requires solving the algebraic equation
$\varphi(t,\vec{x}) + H^{-1} \xi^{0}(t,\vec{x}) = 0$. 
This equation always admits a solution as long as the background spacetime evolves in time, \textit{i.e.}, as long as $H$ does not vanish, which is always satisfied in cosmological applications. 

In the same manner, we can classify all remaining gauge choices. 
For the $(\psi,\alpha)$-sector, the gauge transformation laws are given by~\cite{Izumi:2012qj,Tomonari:2025kgb}
\begin{equation}
\begin{split}
    &\alpha' = \alpha - a^{-1}\xi^{0} \,,\\
    &\psi' = \psi - \dot{\xi}^{0} \,.
\end{split}
\end{equation}
According to the criterion established above, the preferable gauge choice is to fix $\xi^{0}$ such that $\alpha' = 0$. 

For the $(B\,,F)$-sector, the gauge transformation laws are given by~\cite{Izumi:2012qj,Tomonari:2025kgb}
\begin{equation}
\begin{split}
    &B' = B - a^{-1} \xi \,,\\
    &F' = F - (\dot{\xi} - H \xi) \,;
\end{split}
\end{equation}
in this case, the preferable gauge choice is to fix $\xi$ such that $B' = 0$. 

For the $(C_{i}\,,G_{i})$-sector, the gauge transformation laws are given by~\cite{Izumi:2012qj,Tomonari:2025kgb}
\begin{equation}
\begin{split}
    &C'_{i} = C_{i} - (2a)^{-1} \xi^{(v)}_{i} \,,\\
    &G'_{i} = G_{i} - (\dot{\xi}^{(v)}_{i} - H \xi^{(v)}_{i}) \,;
\end{split}
\end{equation}
the preferable gauge choice in this sector is to fix $\xi^{(v)}_{i}$ such that $C'_{i} = 0$. 

Similarly, for the $(C_{i}\,,\tilde{V}_{i})$-sector, the gauge transformation laws are given by~\cite{Izumi:2012qj,Tomonari:2025kgb}
\begin{equation}
\begin{split}
    &C'_{i} = C_{i} - (2a)^{-1} \xi^{(v)}_{i} \,,\\
    &\tilde{V}'_{i} = \tilde{V}_{i} - a^{-1}\epsilon_{ijk} \partial_{j}{\xi^{(v)}_{k}} \,. 
\end{split}
\end{equation}
Again, the preferable gauge choice is to fix $\xi^{(v)}_{i}$ such that $C'_{i} = 0$. 

Summarizing the above discussion, we obtain the following preferable gauge choices:
\begin{enumerate}
    \item Gauge choice I. $\varphi' = 0$, $B' = 0$, $C'_{i} = 0$ (spatially flat gauge); 
    \item Gauge choice II. $\alpha' = 0$, $B' = 0$, $C'_{i} = 0$.
\end{enumerate}
Here, each gauge choice is expressed in terms of perturbation fields rather than the gauge parameters $\xi^{0}$, $\xi$, and $\xi^{(v)}_{i}$.

Finally, we establish a criterion for selecting an appropriate gauge choice from the list given above. 
To this end, we restrict our attention to theories of gravity that satisfy diffeomorphism invariance. 
That is, the DB analysis of such theories confirms that the HDA algebra given in Eq.~\eqref{HDA} is satisfied. 
(See Sec.~\ref{03-01}.)
As indicated in Ref.~\cite{Tomonari:2025kgb} and at the beginning of this subsection, under this assumption the pure spacetime-sector perturbations associated with the diffeomorphism sector do not constitute independent propagating modes. In Type~3, the physical scalar and vector modes are instead associated with the broken boost sector, although in other parametrizations they may appear in mixed combinations of symmetric and antisymmetric parts of vierbein perturbations. 

Therefore, the criterion for gauge choice should be based on whether the internal symmetry is preserved or violated. 
The remaining six perturbative modes, $\alpha$, $\tilde{\sigma}$, $\alpha_{i}$, and $\tilde{V}_{i}$, can propagate only when the internal symmetry described by Eqs.~\eqref{symmetry G} and~\eqref{symmetry T rtimes G} is at least partially broken. 
If this symmetry is preserved, both Gauge choice I and II are admissible. 
On the other hand, if the symmetry is violated, we should select Gauge choice I. 
According to this criterion, Gauge choice I is the most appropriate choice for perturbative analyses of NGR, since NGR does not respect local Lorentz invariance except for Type~6 (TEGR)~\cite{Tomonari:2024ybs,Tomonari:2024lpv}. 

In the next Sec.~\ref{04}, we analyze Type~3 using the \textit{unpreferable} gauge choices $\psi' = 0$ (instead of $\varphi' = 0$), $B' = 0$, and $\tilde{V}'_{i} = 0$ (instead of $C'_{i} = 0$), in order to explicitly scrutinize the issues discussed throughout this subsection. 
The choice $\psi' = 0$ raises doubts regarding the well-definedness of the gauge transformation. 
Furthermore, the choice $\tilde{V}'_{i} = 0$ raises concerns not only about the well-definedness but also about the propagation of modes, due to the violation of local Lorentz invariance. 
A perturbative analysis of NGR around the flat FLRW spacetime employing Gauge choice I can be found in Ref.~\cite{Tomonari:2025kgb}, which adopts the most preferable gauge choice for this theory. 

\paragraph{Remark.}
We emphasize that the discussion in this subsection does not exclude the possibility of other gauge choices beyond Gauge choices I and II. 
For instance, if the integral in Eq.~\eqref{integral of xi0} exists, the $(\varphi\,,\psi)$-sector admits the gauge choice $\psi' = 0$. 
The same property holds for all other sectors. 
This is precisely why we refer to Gauge choice I and II as `\textit{preferable}' gauge choices, rather than merely `\textit{possible}' ones. 
The essential message of this subsection is therefore the following:
\begin{quote}
    \textit{One must choose a gauge that fixes perturbation fields in such a way that the symmetries of the original theory --- revealed, for instance, by DB analysis --- are preserved.}
\end{quote}
The `\textit{preferability}' will particularly be relevant in investigating our theory numerically, and we will address this point further in a sequel paper in detail. 

\subsection{\label{03-03}Counting propagating modes in linear perturbation theory}

We consider gauge-invariant variables associated with the preferable gauge choices, Gauge choices I and II, and give an overview of how propagating modes are counted in linear perturbation theory for a generic theory of gravity.

The gauge-invariant variables are given by~\cite{Tomonari:2025kgb}
\begin{equation}
    \beta = \psi + \varphi - a \dot{\alpha} \,,
    \qquad 
    \gamma = \varphi + \dot{a}\alpha \,,
    \qquad 
    A = F - a \dot{B} \,,
    \qquad 
    B_{i} = G_{i} - 2 a \dot{C}_{i} \,,
    \qquad
    \tilde{W}_{i} = \tilde{V}_{i} - 2 \epsilon_{i j k} \partial_{j}{C_{k}} \,.
\label{gauge invarinant variables}
\end{equation}
In both Gauge choices I and II, $B' = 0$ and $C'_{i} = 0$ hold. 
Thus, we obtain 
\begin{equation}
\begin{split}
    A' = F' (= A) \,,
    \qquad
    B'_{i} = G'_{i} (= B_{i}) \,,
    \qquad
    \tilde{W}'_{i} = \tilde{V}'_{i} (= \tilde{W}_{i}) \,.
\end{split}
\label{gauge invariant variables common in Gauge choice I and II}
\end{equation}
Therefore, if a kinetic term of $F'$, $G'_{i}$, or $\tilde{V}'_{i}$ appears without mutual couplings, each of them can be counted directly as a propagating scalar, vector, or pseudo-vector mode, respectively. 

In the case of scalar modes, we must consider each gauge choice separately.
First, we investigate Gauge choice I, namely $\varphi' = 0$.
Then, the first and second gauge-invariant variables in Eq.~\eqref{gauge invarinant variables} reduce to
\begin{equation}
    \beta' = \psi' - a \dot{\alpha'}(= \beta) \,,
    \qquad 
    \gamma' = \dot{a}\alpha' (= \gamma) \,.
\label{scalar gauge invariant variables in Gauge choice I}
\end{equation}
There are two possible cases in counting propagating modes:
\begin{enumerate}
    \item Case A. The perturbed theory of gravity contains only a kinetic term of $\alpha'$;  
    \item Case B. The perturbed theory of gravity contains not only a kinematic term of $\alpha'$ but also a quadratic term composed of $\psi'$ and/or a time derivative of $\psi'$. 
\end{enumerate}
In Case A, we can count the kinetic term of $\alpha'$ as a propagating scalar mode, since $\dot{\alpha'}\dot{\alpha'}$ contains only the kinetic term of the gauge-invariant variable $\gamma'$, \textit{i.e.}, $(aH)^{-2} \dot{\gamma'}\dot{\gamma'} \in \dot{\alpha'}\dot{\alpha'}$. 
Cases B is more complicated. 
In Case B, there is a potential concern regarding the existence of an Ostrogradsky ghost mode, since $\dot{\psi'}\dot{\psi'}$ contains not only terms proportional to $\dot{\beta'}\dot{\beta'}$ and $\dot{\gamma'}\dot{\gamma'}$, but also a term proportional to $\ddot{\gamma'}\ddot{\gamma'}$.
Furthermore, the terms $\psi'\psi'$ and $\psi'\dot{\psi}'$ contain contributions proportional to $\dot{\gamma}'\dot{\gamma}'$ and to $\ddot{\gamma}'$, respectively.
In the latter case, there is again a concern about the presence of an Ostrogradsky ghost mode since $\gamma'$ is gauge-invariant (\textit{i.e.} physical) variable.
Therefore, it is necessary to investigate a degeneracy condition to eliminate such ghost modes according to Refs.~\cite{Ostrogradsky:1850fid,Pons:1988tj,Woodard:2015zca,Motohashi:2016ftl}. 
This condition will trap at least one of $\gamma$ and $\beta$ by this condition, thereby the total number of propagating modes is in turn up to eight. 
As long as we pay careful attention to this point, if the mode $\alpha'$ decouples from the other modes, its kinetic term can also be counted as a propagating scalar mode. 
In our previous work~\cite{Tomonari:2025kgb}, we applied Gauge choice I in the perturbative analysis, and Case A was concluded. 
Thus, we correctly identified the propagation of the scalar mode, and the absence of an Ostrogradsky ghost mode is guaranteed in principle.

Second, we consider Gauge choice II, namely $\alpha' = 0$.
Then, the first and second gauge-invariant variables in Eq.~\eqref{gauge invarinant variables} reduce to
\begin{equation}
    \beta' = \psi' + \varphi' (= \beta) \,,
    \qquad 
    \gamma' = \varphi' (= \gamma) \,.
\label{scalar gauge invariant variables in Gauge choice II}
\end{equation}
There are two possible cases in counting propagating modes: 
\begin{enumerate}
    \item Case A. The perturbed theory of gravity contains only the kinetic term of $\varphi'$; 
    \item Case B. The perturbed theory of gravity contains the kinetic terms not only of $\varphi'$ but also of $\psi'$. 
\end{enumerate}
In Case A, the diffeomorphism invariance prohibits this mode from propagating; 
this case cannot be realized under this invariance.
In Case B, the term $\dot{\psi'}\dot{\psi'}$ contains two kinetic terms proportional to $\dot{\beta'}\dot{\beta'}$ and $\dot{\gamma'}\dot{\gamma'}$, respectively, and no higher-order time derivative appears. 
Therefore, Ostrogradsky instability is absent in principle, although the diffeomorphism invariance again prevents these modes from propagating.

We summarize the investigation of this section in Table~\ref{Table:PropagatingModesImPreferableGaugeChoice}. 
In the preferable gauge choices, all perturbation fields \textit{can} be expressed solely in terms of gauge-invariant variables.
By contrast, in other gauge choices, the perturbed theory \textit{cannot} be expressed solely in terms of gauge-invariant variables. 
This result can be applied not only to NGR but also to other MAG theories with the Weitzenb\"{o}ck gauge. 
\begin{table}[ht!]
    \centering
    \renewcommand{\arraystretch}{1.5}
        \caption{
    \textbf{Possible propagating modes in Gauge choices I and II.}\\
    We assume that diffeomorphism invariance is preserved.
    The symbols $\times$ and $\checkmark$ indicate that local Lorentz invariance is violated and preserved at least in the boost sector, respectively. 
    All perturbation fields are expressed in terms of gauge-invariant variables.
    In Case A for Gauge choice I, the propagating scalar mode $\gamma$ can be counted via $\alpha$, in the perturbed Lagrangian, even though the variable itself is not gauge invariant. 
    }
    \begin{tabular}{ c || c || c }
    	Gauge choice 
        &
        Gauge choice I (Boost invariance: $\times$)
        & 
        Gauge choice II (Boost invariance: $\checkmark$)
        \\ \hline \hline 
    	Case 
        &
        \begin{tabular}{c|c}
            Case A & Case B   
        \end{tabular}
        &
        \begin{tabular}{c|c}
            Case A & Case B 
        \end{tabular}
        \\ \hline 
        \begin{tabular}{c}
            Possible propagating \\
            tensor mode  
        \end{tabular}
        &
        \begin{tabular}{c|c}
            $h_{ij}$ & $h_{ij}$
        \end{tabular}
        &
        \begin{tabular}{c|c}
            $h_{ij}$ & $h_{ij}$
        \end{tabular}
        \\ \hline 
        \begin{tabular}{c}
            Possible propagating \\
            (pseudo-)vector mode  
        \end{tabular}
        &
        \begin{tabular}{c}
            \begin{tabular}{c|c}
                $\alpha_{i}\,;\,\tilde{V}_{i} (=\tilde{W}_{i})$ & $\,,\alpha_{i}\,;\,\tilde{V}_{i} (=\tilde{W}_{i})$
            \end{tabular}
            \\ \hline 
            N.B. If the rotation invariance is violated, \\
            $\tilde{V}_{i} (=\tilde{W}_{i})$ can propagate in both cases
        \end{tabular}
        &
        \begin{tabular}{c}
            \begin{tabular}{c|c}
            $\tilde{V}_{i} (=\tilde{W}_{i})$ & $\tilde{V}_{i} (=\tilde{W}_{i})$ 
            \end{tabular} 
            \\ \hline
            N.B. If the rotation invariance is violated, \\
            $\tilde{V}_{i} (=\tilde{W}_{i})$ can propagate in both cases
        \end{tabular}
        \\ \hline 
        \begin{tabular}{c}
            Possible propagating \\
            (pseudo-)scalar mode  
        \end{tabular}
        &
        \begin{tabular}{c}
            \begin{tabular}{c|c}
                $\gamma\,;\,\tilde{\sigma}$ & 
                \begin{tabular}{c}
                    $\mathrm{either\,}\gamma\mathrm{\,or\,}\beta\,;\,\tilde{\sigma}$ \\
                    \hline 
                    N.B. Ostrogradsky instability exist. 
                \end{tabular}
            \end{tabular}
            \\ \hline
            N.B. If the rotation invariance is violated, \\
            $\tilde{\sigma}$ can propagate in both cases
        \end{tabular}
        &
        \begin{tabular}{c}
            \begin{tabular}{c|c}
            $\tilde{\sigma}$ & $\tilde{\sigma}$
            \end{tabular}
            \\ \hline 
            N.B. If the rotation invariance is violated, \\
            $\tilde{\sigma}$ can propagate in both cases
        \end{tabular}
        \\ \hline \hline
    \end{tabular}
    \label{Table:PropagatingModesImPreferableGaugeChoice}
\end{table}

\section{\label{04} Perturbation of Type~3 revisited with an \textit{un-}preferable gauge choice}

Taking into account the discussion in Secs.~\ref{02} and~\ref{03-01}, Gauge choice I is preferable for the perturbative analysis of Type~3, owing to the presence of diffeomorphism and rotation symmetries. 
In our previous work~\cite{Tomonari:2025kgb}, we performed the perturbative analysis using Gauge choice I and found that Case A is concluded. 
We identified the existence of tensor, vector, and scalar modes in Type~3. 
In particular, the vector and scalar modes are attributed to the violation of symmetry in the boost sector of the local Lorentz invariance. 

In the following subsections, as mentioned at the end of Sec.~\ref{03-02}, we examine the perturbative analysis in the \textit{un-}preferable gauge choice $\psi' = 0$ (instead of $\varphi' = 0$) and $\tilde{V}'_{i} = 0$ (instead of $C'_{i} = 0$), and investigate its \textit{in-}validity for counting propagating modes. 
Throughout this section, we use Cadabra~\cite{Peeters:2007wn} for all computations. 

\subsection{\label{04-01} Background equations and tensor perturbation} 

We can derive the background equations of NGR from Eq.~\eqref{field equation of NGR} as follows~\cite{Tomonari:2025kgb}: 
\begin{equation}
\begin{split}
    &
    - 3 (2c_{1}-c_{2}+3c_{3})H^{2} 
    = 
    - \frac{1}{2} \dot{\Phi}_{0} \dot{\Phi}_{0} - V_{0} 
    \,,
    \\
    & 
    - ( 2 c_{1} - c_{2} + 3 c_{3} )( 3 H^{2} + 2 \dot{H} ) 
    = 
    - V_{0} + \frac{1}{2} \dot{\Phi}_{0} \dot{\Phi}_{0} 
    \,,
\end{split}
\label{bcgd eqs of NGR with test matter field}
\end{equation}
where $\Phi_{0}$ is the background scalar field of a matter source $\Phi$ and $V_{0}=V(\Phi_{0})$. 
The dot `$\,\cdot\,$' over on each object denotes the time derivative. 
We perturbed $\Phi$ by $\Phi = \Phi_{0} + \delta\Phi$. 
By setting $c_{1} = - 1/4$, $c_{2} = 1/2$, $c_{3} = 1$, we recover the field equations of TEGR (Type~6). 
The coefficient $2 c_{1} - c_{2} + 3 c_{3}$ controls the FLRW background coupling with matter fields. 
In addition, by combining the above equations, we obtain an equation describing the time evolution of the Hubble parameter~\cite{Tomonari:2025kgb}:
\begin{equation}
    - 2 ( 2 c_{1} - c_{2} + 3 c_{3} ) \dot{H} 
    = 
    \dot{\Phi}_{0} \dot{\Phi}_{0} 
    \,.
\label{simplified bcgd eqs of NGR with test matter filed}
\end{equation}

The background field equation for the scalar field around the background spacetime is given by~\cite{Tomonari:2025kgb}
\begin{equation}
    - \ddot{\Phi}_{0} - 3 H \dot{\Phi}_{0} - V'_{0} = 0 \,, 
\label{bcgd eq of test matter field}
\end{equation}
where $V'_{0}$ denotes the first derivative of $V$ with respect to $\Phi$ evaluated at $\Phi_{0}$: $V'_{0} = V'(\Phi_{0})$. 
By combining Eq.~\eqref{simplified bcgd eqs of NGR with test matter filed} with Eq.~\eqref{bcgd eq of test matter field}, we can solve for the Hubble parameter and the background scalar field. 

Based on this set-up, the tensor perturbation is derived as follows~\cite{Tomonari:2025kgb}: 
\begin{equation}
\begin{split}
    & \mathcal{L}^\mathrm{(TP\,;\,2nd\,order)}_\mathrm{total} 
    =
    \\ 
    &
    - a^{3}\,\left(2\,c_{1}- c_{2}\right)\,\delta^{ik}\,\delta^{jl}\,\dot{h}_{ij}\,\dot{h}_{kl} 
    + 2 \left(2\,c_{1} - c_{2} + 3\,c_{3}\right)\,a^{3}\,H\,\delta^{ik}\,\delta^{jl}\,\dot{h}_{ij}\,h_{kl} \\
    & 
    - \frac{1}{2}\,a^{3}\,H^{2}\,\delta^{ik}\,\delta^{jl}\,h_{ij}\,h_{kl} 
    + a\,\left(2\,c_{1} + c_{2}\right)\,
    \delta^{il}\,\delta^{jm}\,\delta^{kn}\,\partial_{i}h_{jk}\,\partial_{l}h_{mn} 
    - a\,\left(2\,c_{1} + c_{2}\right) \,\delta^{im}\,\delta^{jn}\,\delta^{kl}\,\partial_{i}h_{jk}\,\partial_{l}h_{mn} 
    \\
    & 
    + \frac{1}{2} a^{3} \dot{\delta\Phi} \dot{\delta\Phi} - \frac{1}{2} a \delta^{i j} \partial_{i}{\delta\Phi} \partial_{j}{\delta\Phi} 
    - \frac{1}{2} a^{3} V''_{0} \delta\Phi \delta\Phi 
    \,.
\end{split}
\label{TPs of NGR in 2nd-order}
\end{equation} 
This perturbative expansion is written in terms of gauge invariant variable. 

\subsection{\label{04-02} Scalar perturbation with an \textit{un-}preferable versus preferable gauge} 

The vierbein and metric components are expanded as follows: 
\begin{equation}
\begin{split}
    e^{0}{}_{\mu}dx^{\mu} 
    &= 
    dt
    +
    a\partial_{i}\alpha dx^{i}
    \,,\\
    e^{a}{}_{\mu}dx^{\mu} 
    &=
    \delta^{ai}\partial_{i}F dt 
    + 
    a\left(1 - \varphi\right)\delta^{a}{}_{i} dx^{i}
    \,,\\
    e_{0}{}^{\mu}\frac{\partial}{\partial x^{\mu}} 
    &=
    \left( 1 + \delta^{ij} \partial_{i}{F} \partial_{j}{\alpha} \right) \frac{\partial}{\partial t} 
    + 
    \left( - a^{-1} (1 + \varphi) \delta^{ij} \partial_{j}{F} \right) \frac{\partial}{\partial x^{i}}
    \,,\\
    e_{a}{}^{\mu}\frac{\partial}{\partial x^{\mu}} 
    &= 
    \left( - ( 1 + \varphi ) \delta_{a}{}^{i} \partial_{i}{\alpha} \right) \frac{\partial}{\partial t} 
    + 
    \left( a^{-1} \delta_{a}{}^{i} \left( 1 + \varphi + \varphi^{2} \right)
    + a^{-1} \delta_{a}{}^{k} \delta^{ij} \partial_{k}{\alpha} \partial_{j}{F} \right) \frac{\partial}{\partial x^{i}}
    \,,\\
    g_{\mu\nu}\,dx^{\mu}dx^{\nu} 
    &= \left[
    -1 + \delta^{i j} \partial_{i}{F} \partial_{j}{F}
    \right] dt dt 
    + 
    2 a \left[ 
    - \partial_{i}{\alpha} + (1 - \varphi) \partial_{i}{F}
    \right] dt dx^{i} 
    \\
    & 
    + a^{2} \left[
    \delta_{i j} \left( 1 - 2 \varphi +  \varphi^{2} \right) 
    - \partial_{i}{\alpha} \partial_{j}{\alpha}
    \right] dx^{i} dx^{j}
    \,,\\
    g^{\mu\nu}\,\frac{\partial}{\partial x^{\mu}}\frac{\partial}{\partial x^{\nu}} 
    &= 
    \left[
    - 1 - 2 \delta^{i j} \partial_{i}{F} \partial_{j}{\alpha} + \delta^{i j} \partial_{i}{\alpha} \partial_{j}{\alpha}
    \right] \partial_{0} \partial_{0} 
    + 2 a^{-1} \delta^{i j} \left[
    ( 1 + \varphi ) \partial_{j}{F}
    - ( 1 + 2 \varphi ) \partial_{j}{\alpha} 
    \right] \partial_{0} \partial_{i} 
    \\
    &
    + a^{-2} \left[
    \delta^{i j} ( 1 + 2 \varphi + 3 \varphi^{2} ) 
    + 2 \delta^{i k} \delta^{j l} \partial_{k}{F} \partial_{l}{\alpha} 
    - \delta^{i k} \delta^{j l} \partial_{k}{F} \partial_{l}{F}
    \right] \partial_{i} \partial_{j}
    \,,\\
    &\theta 
    = 
    a^{3} \left(
    1 - 3 \varphi + 3 \varphi^{2} - \delta^{ij} \partial_{i}{F} \partial_{j}{\alpha} 
    \right)
    \,. 
\end{split}
\label{vierbein and metric in SPs}
\end{equation}

The second-order action for scalar perturbations is calculated as follows: 
\begin{equation}
\begin{split}
    &\mathcal{L}^\mathrm{(SP\,;\,2nd\,order)}_\mathrm{total} = 
    \\
    &
    + ( 2 c_{1} - c_{2} + c_{3} ) a^{3} \delta^{i j} \partial_{i}{\dot{\alpha}} \partial_{j}{\dot{\alpha}} 
    + 2 ( 2 c_{1} - c_{2} + 3 c_{3} ) H a^{3} \delta^{i j} \partial_{i}{\alpha} \partial_{j}\dot{\alpha} 
    + 3 ( 2 c_{1} - c_{2} + 3 c_{3} ) H^{2} a^{3} \delta^{i j} \partial_{i}{\alpha} \partial_{j}{\alpha} 
    \\
    &
    + 4 c_{3} a^{2} \delta^{i j} \partial_{i}{\varphi} \partial_{j}\dot{\alpha} 
    - 3 ( 2 c_{1} - c_{2} + 3 c_{3} ) H^{2} a^{3} \delta^{i j} \partial_{i}{F} \partial_{j}{\alpha} 
    + 4 ( 2 c_{1} - c_{2} + 3 c_{3} ) H a^{2} \delta^{i j} \partial_{i}{\alpha} \partial_{j}{\varphi} 
    \\
    &
    - \frac{1}{2} a^{3} \dot{\Phi}_{0} \dot{\Phi}_{0} \delta^{i j} \partial_{i}{\alpha} \partial_{j}{\alpha} 
    + a^{3} V_{0} \delta^{i j} \partial_{i}{F} \partial_{j}{\alpha} 
    + a^{2} \dot{\Phi}_{0} \delta^{i j} \partial_{i}{\alpha} \partial_{j}{\delta\Phi}  
    + \frac{1}{2} a^{3} \dot{\Phi}_{0} \dot{\Phi}_{0} \delta^{i j} \partial_{i}{F} \partial_{j}{\alpha} 
    \\
    &
    - 3 ( 2 c_{1} - c_{2} + 3 c_{3} ) a^{3} \dot{\varphi} \dot{\varphi} 
    - 12 ( 2 c_{1} - c_{2} + 3 c_{3} ) H a^{3} \varphi \dot{\varphi} 
    - 2 ( 2 c_{1} - c_{2} + 3 c_{3} ) a^{2} \dot{\varphi} \delta^{i j} \partial_{i} \partial_{j}{F} 
    + 3 a^{3} \dot{\Phi}_{0} \dot{\varphi} \delta\Phi 
    \\
    &
    + 2 ( 2 c_{1} - c_{2} + 2 c_{3} ) a \delta^{i j} \partial_{i}{\varphi} \partial_{j}{\varphi} 
    - a^{2} \dot{\Phi}_{0} \delta^{i j}  \partial_{i}{F} \partial_{j}{\delta\Phi} 
    - ( 2 c_{1} - c_{2} + c_{3} ) a \delta^{i k} \delta^{j l} \partial_{i}\partial_{j}{F} \partial_{k}\partial_{l}{F} 
    \\
    &
    + \frac{1}{2} a^{3} \dot{\delta\Phi} \dot{\delta\Phi}  
    - \frac{1}{2} a \delta^{i j} \partial_{i}{\delta\Phi} \partial_{j}{\delta\Phi} 
    - \frac{1}{2} a^{3} V''_{0} \delta\Phi \delta\Phi 
    \,,
\end{split}
\label{SP of NGR}
\end{equation}
where the background equations, Eqs.~\eqref{bcgd eqs of NGR with test matter field} and~\eqref{bcgd eq of test matter field}, have been used. 
$V''_{0}$ denotes the second derivative of $V$ with respect to $\Phi$ evaluated at $\Phi_{0}$: $V''_{0} = V''(\Phi_{0})$. 
By varying the action with respect to $F$ and using Eq.~\eqref{bcgd eqs of NGR with test matter field}, we obtain a constraint equation.
Substituting this constraint back into Eq.~\eqref{SP of NGR}, we arrive at
\begin{equation}
\begin{split}
    &\therefore\qquad 
    \mathcal{L}^\mathrm{(SP\,;\,2nd\,order)}_\mathrm{total} = 
    \\
    &
    + ( 2 c_{1} - c_{2} + c_{3} ) a^{3} \delta^{i j} \partial_{i}{\dot{\alpha}} \partial_{j}{\dot{\alpha}} 
    + 2 ( 2 c_{1} - c_{2} + 3 c_{3} ) H a^{3} \delta^{i j} \partial_{i}{\alpha} \partial_{j}\dot{\alpha} 
    + 3 ( 2 c_{1} - c_{2} + 3 c_{3} ) H^{2} a^{3} \delta^{i j} \partial_{i}{\alpha} \partial_{j}{\alpha} 
    \\
    &
    + 4 c_{3} a^{2} \delta^{i j} \partial_{i}{\varphi} \partial_{j}\dot{\alpha} 
    + 4 ( 2 c_{1} - c_{2} + 3 c_{3} ) H a^{2} \delta^{i j} \partial_{i}{\alpha} \partial_{j}{\varphi} 
    - \frac{1}{2} a^{3} \dot{\Phi}_{0} \dot{\Phi}_{0} \delta^{i j} \partial_{i}{\alpha} \partial_{j}{\alpha} 
    + a^{2} \dot{\Phi}_{0} \delta^{i j} \partial_{i}{\alpha} \partial_{j}{\delta\Phi}  
    \\
    &
    - 3 ( 2 c_{1} - c_{2} + 3 c_{3} ) a^{3} \dot{\varphi} \dot{\varphi} 
    - 12 ( 2 c_{1} - c_{2} + 3 c_{3} ) H a^{3} \varphi \dot{\varphi} 
    + 3 a^{3} \dot{\Phi}_{0} \dot{\varphi} \delta\Phi 
    + 2 ( 2 c_{1} - c_{2} + 2 c_{3} ) a \delta^{i j} \partial_{i}{\varphi} \partial_{j}{\varphi} 
    \\
    &
    + \frac{1}{2} a^{3} \dot{\delta\Phi} \dot{\delta\Phi}  
    - \frac{1}{2} a \delta^{i j} \partial_{i}{\delta\Phi} \partial_{j}{\delta\Phi} 
    - \frac{1}{2} a^{3} V''_{0} \delta\Phi \delta\Phi 
    \,.
\end{split}
\label{SP of NGR with constraint wrt F}
\end{equation}
Here, we have integrated out decoupled terms that do not contain any time derivatives.
At first glance, this Lagrangian appears to contain two propagating scalar modes.
However, it should be emphasized that the action is not written solely in terms of gauge-invariant variables.
From a general perspective, Type~3 NGR admits only a single scalar mode, which originates from the violation of local Lorentz symmetry in the boost sector (see Sec.~\ref{04-02}).
The variable $\alpha$ is associated with this symmetry breaking, whereas $\varphi$ does not correspond to a physical scalar degree of freedom.

To make this point clear, we consider the case of Type~6 (TEGR).
Imposing the Type~6 condition, $2c_{1} - c_{2} + c_{3} = 0$, on Eq.~\eqref{SP of NGR with constraint wrt F}, we obtain
\begin{equation}
\begin{split}
    &\therefore\qquad 
    \mathcal{L}^\mathrm{(SP\,;\,2nd\,order)}_\mathrm{Type\,6} = 
    \\
    &
    + 2 ( 2 c_{1} - c_{2} + 3 c_{3} ) H a^{3} \delta^{i j} \partial_{i}{\alpha} \partial_{j}\dot{\alpha} 
    + 3 ( 2 c_{1} - c_{2} + 3 c_{3} ) H^{2} a^{3} \delta^{i j} \partial_{i}{\alpha} \partial_{j}{\alpha} 
    \\
    &
    + 4 c_{3} a^{2} \delta^{i j} \partial_{i}{\varphi} \partial_{j}\dot{\alpha} 
    + 4 ( 2 c_{1} - c_{2} + 3 c_{3} ) H a^{2} \delta^{i j} \partial_{i}{\alpha} \partial_{j}{\varphi} 
    - \frac{1}{2} a^{3} \dot{\Phi}_{0} \dot{\Phi}_{0} \delta^{i j} \partial_{i}{\alpha} \partial_{j}{\alpha} 
    + a^{2} \dot{\Phi}_{0} \delta^{i j} \partial_{i}{\alpha} \partial_{j}{\delta\Phi}  
    \\
    &
    - 3 ( 2 c_{1} - c_{2} + 3 c_{3} ) a^{3} \dot{\varphi} \dot{\varphi} 
    - 12 ( 2 c_{1} - c_{2} + 3 c_{3} ) H a^{3} \varphi \dot{\varphi} 
    + 3 a^{3} \dot{\Phi}_{0} \dot{\varphi} \delta\Phi 
    + 2 ( 2 c_{1} - c_{2} + 2 c_{3} ) a \delta^{i j} \partial_{i}{\varphi} \partial_{j}{\varphi} 
    \\
    &
    + \frac{1}{2} a^{3} \dot{\delta\Phi} \dot{\delta\Phi}  
    - \frac{1}{2} a \delta^{i j} \partial_{i}{\delta\Phi} \partial_{j}{\delta\Phi} 
    - \frac{1}{2} a^{3} V''_{0} \delta\Phi \delta\Phi 
    \,.
\end{split}
\label{SP of Type 6 NGR}
\end{equation}
Repeating the same procedure, we obtain
\begin{equation}
\begin{split}
    &\therefore\qquad 
    \mathcal{L}^\mathrm{(SP\,;\,2nd\,order)}_\mathrm{Type\,6} = 
    \\
    &
    - 3 ( 2 c_{1} - c_{2} + 3 c_{3} ) a^{3} \dot{\varphi} \dot{\varphi}  
    - 12 ( 2 c_{1} - c_{2} + 3 c_{3} ) H a^{3} \varphi \dot{\varphi} 
    + 3 a^{3} \dot{\Phi}_{0} \dot{\varphi} \delta\Phi 
    + 2 ( 2 c_{1} - c_{2} + 2 c_{3} ) a \delta^{i j} \partial_{i}{\varphi} \partial_{j}{\varphi} 
    \\
    &
    + \frac{1}{2} a^{3} \dot{\delta\Phi} \dot{\delta\Phi}  
    - \frac{1}{2} a \delta^{i j} \partial_{i}{\delta\Phi} \partial_{j}{\delta\Phi} 
    - \frac{1}{2} a^{3} V''_{0} \delta\Phi \delta\Phi 
    \,.
\end{split}
\label{SP of Type 6 NGR with constraint wrt alpha}
\end{equation}
where we have integrated out the decoupled terms that do not contain any time derivatives.
Since Type~6 (TEGR) is equivalent to GR, this theory must contain only the tensor propagating mode.

This issue arises from an inappropriate gauge choice.
In fact, if we adopt Gauge choice I, we obtain~\cite{Tomonari:2025kgb}
\begin{equation}
\begin{split}
    & \mathcal{L}^\mathrm{(SP\,;\,2nd\,order)}_\mathrm{Type\,3\,in\,gauge\text{-}dependent\,form}  
    \\
    &= 
    ( 2 c_{1} - c_{2} + c_{3} ) a^{3} \delta^{ij} \partial_{i}{\dot{\alpha}} \partial_{j}{\dot{\alpha}}
    + 4 ( 2 c_{1} - c_{2} + 3 c_{3} ) a^{3} H \delta^{ij} \partial_{i}{\dot{\alpha}} \partial_{j}{\alpha} 
    - 3 ( c_{2} - 3 c_{3} ) a^{3} H^{2} \delta^{ij} \partial_{i}{\alpha} \partial_{j}{\alpha} 
    \\
    &\qquad
    + \frac{1}{2} a^{3} \dot{\Phi}_{0} \dot{\Phi}_{0} \delta^{ij} \partial_{i}{\alpha} \partial_{j}{\alpha}
    + 2 a^{2} \dot{\Phi}_{0} \delta^{ij} \partial_{i}{\alpha} \partial_{j}{\delta\Phi}
    \\
    &\qquad
    + \frac{1}{2} a^{3} \dot{\delta\Phi} \dot{\delta\Phi}
    - \frac{1}{2} a \delta^{ij} \partial_{i}{\delta\Phi} \partial_{j}{\delta\Phi}
    - \frac{1}{2} a^{3} V''_{0} \delta\Phi \delta\Phi
    \,,
\end{split}
\label{SPs in Type 3 with gauge choice I}
\end{equation}
where we have integrated out the decoupled terms that do not contain any time derivatives.
This perturbed Lagrangian is expressed solely in terms of gauge-invariant variables, except for $\alpha$.
However, as discussed in Sec.~\ref{03-03}, the kinetic term of $\alpha$ is in one-to-one correspondence with that of the gauge-invariant variable $\gamma$.
Since this Lagrangian is classified as Case A of Gauge choice I in Table~\ref{Table:PropagatingModesImPreferableGaugeChoice}, no Ostrogradsky instability arises.
Therefore, we can consistently count this mode as a propagating scalar degree of freedom, as was done in our previous work~\cite{Tomonari:2025kgb}.

We can translate Eq.~\eqref{SPs in Type 3 with gauge choice I} in terms of the gauge-invariant variable $\gamma$ as follows:
\begin{equation}
\begin{split}
    &\mathcal{L}^\mathrm{(SP\,;\,2nd\,order)}_\mathrm{Type\,3} \\
    &
    = 
    ( 2 c_{1} - c_{2} + c_{3} ) a H^{-2} \delta^{i j} \partial_{i}{\dot{\gamma}} \partial_{j}{\dot{\gamma}} 
    - 2 ( 2 c_{1} - c_{2} + c_{3} ) c a^{2} H^{-1} \delta^{i j} \partial_{i}{\gamma} \partial_{j}{\dot{\gamma}} 
    - 4 ( 2 c_{1} - c_{2} + 3 c_{3} ) a H^{-1} \delta^{i j}  \partial_{i}{\gamma} \partial_{j}{\dot{\gamma}} 
    \\
    & 
    + ( 2 c_{1} - c_{2} + c_{3} ) c^{2} a^{3} \delta^{i j} \partial_{i}{\gamma} \partial_{j}{\gamma} 
    + 4 ( 2 c_{1} - c_{2} + 3 c_{3} ) c a^{2} \delta^{i j} \partial_{i}{\gamma} \partial_{j}{\gamma} 
    - 3 ( c_{2} - 3 c_{3} ) a \delta^{i j} \partial_{i}{\gamma} \partial_{j}{\gamma}
    \\
    &
    + \frac{1}{2}aH^{-2}\dot{\Phi}_{0}{}\dot{\Phi}_{0}{}\delta^{ij}\partial_{i}{\gamma}\partial_{j}{\gamma} 
    + 2aH^{-1}\dot{\phi}_{0}\delta^{ij}\partial_{i}{\gamma}\partial_{j}{\delta\Phi}
    \\
    &
    + \frac{1}{2} a^{3} \dot{\delta\Phi} \dot{\delta\Phi} 
    - \frac{1}{2} a \delta^{ij} \partial_{i}{\delta\Phi} \partial_{j}{\delta\Phi} 
    - \frac{1}{2} a^{3} V''_{0} \delta\Phi \delta\Phi 
    \,,
\end{split}
\label{SPs in Type 3 with gauge choice I in gauge invariant variable}
\end{equation}
where we set 
\begin{equation}
    c 
    = 
    - \frac{1}{a} \left[
    1 - \frac{1}{H^{2} (2 c_{1} - c_{2} + 3 c_{3} )} \left( \frac{1}{2} \dot{\Phi}^{2}_{0} - V_{0} \right)
    \right]
    =
    - \frac{2}{a}\left(2 + \frac{\dot{H}}{H^{2}}\right)
    \,.
\label{parameter c in SPs in Type 3 with gauge choice I in gauge invariant variable}
\end{equation}
where we have used the background equations given in Eq.~\eqref{bcgd eqs of NGR with test matter field}. 

The determinant of the kinetic matrix of this type is 
\begin{equation}
\begin{split}
    &\mathcal{K}_\mathrm{SP} 
    := 
    \mathrm{det}\left(\frac{\delta^{2}\mathcal{L}^\mathrm{(SP\,;\,2nd\,order)}_\mathrm{Type\,3}(t\,,\vec{z})}{\delta\dot{X}_{i}(t\,,\vec{x})\delta\dot{X}_{j}(t\,,\vec{y})}\right) 
    = 
    \frac{1}{2} a^{4} H^{-2} ( 2 c_{1} - c_{2} + c_{3} ) 
    \Delta_\mathrm{SP}(\vec{x}\,,\vec{y}\,,\vec{z})
    \,,\\
    &\Delta_\mathrm{SP} (\vec{x}\,,\vec{y}\,,\vec{z}) 
    = 
    \delta^{i j} \partial^{(z)}_{i}{\delta^{(3)}(\vec{z} - \vec{x})} \partial^{(z)}_{j}{\delta^{(3)}(\vec{z} - \vec{y})}\cdot\delta^{(3)}(\vec{z} - \vec{x})\delta^{(3)}(\vec{z} - \vec{y})
    \,,
\end{split}
\label{det of kinetic matrix in SPs}
\end{equation}
where $X_{i} = \{\gamma(t\,,\vec{x})\,,\delta\Phi(t\,,\vec{x})\}$ and $\partial^{(z)}_{i}$ denotes the partial derivative with respect to $\vec{z}$. 
Thus, the ghost free condition is determined only by the kinetic term. 
We remark that one can verify that the kinetic matrix considered here coincides with the non-degenerate block of the kinetic matrix calculated under the relation of Eq.~\eqref{scalar gauge invariant variables in Gauge choice I} from the unreduced Lagrangian given in Ref.~\cite{Tomonari:2025kgb}. 

\subsection{\label{04-03} Vector perturbation with an \textit{un-}preferable versus preferable gauge} 

The vierbein and metric components are expanded as follows: 
\begin{equation}
\begin{split}
    e^{0}{}_{\mu}dx^{\mu} 
    &= 
    dt
    +
    a \alpha_{i} dx^{i}
    \,,\\
    e^{a}{}_{\mu}dx^{\mu} 
    &= 
    \delta^{aj} G_{j} dt 
    + 
    a \left[
    \delta^{a}{}_{i} 
    + 2 \delta^{aj} \partial_{(i}C_{j)} 
    \right] dx^{i}
    \,,\\
    e_{0}{}^{\mu}\frac{\partial}{\partial x^{\mu}} 
    &= 
    \left[
    1 + \delta^{ij} G_{i} \alpha_{j} 
    \right] \frac{\partial}{\partial t} 
    +
    \left[
    - a^{-1} \delta^{ij} G_{j} 
    - 2 a^{-1} \delta^{ij} \delta^{kl} G_{k} \partial_{(l}C_{j)}
    \right] \frac{\partial}{\partial x^{i}}
    \,,\\
    e_{a}{}^{\mu}\frac{\partial}{\partial x^{\mu}} 
    &= 
    \left[
    - \delta^{i}{}_{a} \alpha_{i} 
    + \delta^{i}{}_{a} \alpha_{i} \cdot 2 \delta^{jk} \partial_{(j}C_{k)} 
    \right] \frac{\partial}{\partial t} \\
    &\qquad
    + \left[
    + a^{-1} \delta^{i}{}_{a} 
    - 2 a^{-1} \delta^{j}{}_{a} \delta^{ik} \partial_{(j}C_{k)} 
    + a^{-1} \delta^{j}{}_{a} \delta^{ik} \alpha_{j} G_{k} 
    + 4 a^{-1} \delta^{m}{}_{a} \delta^{kl} \partial_{(m}C_{k)} \delta^{ij} \partial_{(l}C_{j)}
    \right] \frac{\partial}{\partial x^{i}}
    \,,\\
    g_{\mu\nu}\,dx^{\mu}dx^{\nu} 
    &= 
    \left[
    - 1 
    + \delta^{i j} G_{i} G_{j}
    \right] dt dt
    +
    2 a \left[
    G_{i} 
    - \alpha_{i} 
    + \delta^{j k} C_{i j} G_{k} 
    \right] dt dx^{i} 
    + 
    a^{2} \left[
    + \delta_{i j}
    - \alpha_{i} \alpha_{j} 
    + 2 C_{i j} 
    + \delta^{k l} C_{i k} C_{j l}
    \right] dx^{i} dx^{j}
    \,,\\
    g^{\mu\nu}\,\frac{\partial}{\partial x^{\mu}}\frac{\partial}{\partial x^{\nu}} 
    &= 
    \left[
    - 1 
    - 2 \delta^{i j} G_{i} \alpha_{j} 
    + \delta^{i j} \alpha_{i} \alpha_{j} 
    \right] \frac{\partial}{\partial t} \frac{\partial}{\partial t} 
    + 
    2 a^{-1} \left[
    \delta^{i j} G_{j} 
    - \delta^{i j} \alpha_{j} 
    + \delta^{l j} \delta^{i k} C_{j k} \alpha_{l}  
    + \delta^{l j} \delta^{i k} C_{j k} G_{l}  
    \right] \frac{\partial}{\partial t} \frac{\partial}{\partial x^{i}} \\
    &
    +
    a^{2} \left[
    \delta^{i j} 
    - 2 \delta^{i k} \delta^{j l} C_{k l} 
    - \delta^{i j} \delta^{j l} G_{j} G_{l} 
    + 2 \delta^{i l} \delta^{j k} G_{k} \alpha_{l} 
    + 3 \delta^{l n} \delta^{i m} \delta^{j k} C_{k l} C_{m n} 
    \right] \frac{\partial}{\partial x^{i}} \frac{\partial}{\partial x^{j}}
    \,,\\
    &\theta 
    = a^{3} \left[
    1 - \delta^{i j} \alpha_{i} G_{j} + \delta^{i j} \delta^{k l} C_{i k} C_{j l} 
    \right]
    \,. 
\end{split}
\label{vierbein and metric in SPs}
\end{equation}

The second-order perturbation is calculated as follows: 
\begin{equation}
\begin{split}
    &\mathcal{L}^\mathrm{(VP\,;\,2nd\,order)}_\mathrm{total} \\
    &= 
    + ( 2 c_{1} - c_{2} + c_{3} ) a^{3} \delta^{i j} \dot{\alpha}_{i} \dot{\alpha}_{j} 
    + 2 ( 2 c_{1} - c_{2} + 3 c_{3} ) H a^{3} \delta^{i j} \alpha_{i} \dot{\alpha}_{j} 
    - 2 c_{1} a^{-1} \delta^{i j} \delta^{k l} \partial_{i}{\alpha_{k}} \partial_{j}{\alpha_{l}} 
    + 3 ( 2 c_{1} - c_{2} + 3 c_{3} ) H^{2} a^{3} \delta^{i j} \alpha_{i} \alpha_{j} \\
    &
    + 2 c_{3} a^{2} \delta^{i j} \delta^{k l} \partial_{i}{C_{j k}} \dot{\alpha}_{l} 
    - 2 c_{2} a^{-1} \delta^{i j} \delta^{k l} \partial_{i}{G_{k}} \partial_{j}{\alpha_{l}}
    + 2 ( 2 c_{1} - c_{2} + 3 c_{3} ) H a^{2} \delta^{i j} \delta^{k l} \alpha_{i} \partial_{k}{C_{j l}} 
    - 3 ( 2 c_{1} - c_{2} + 3 c_{3} ) H^{2} a^{3} \delta^{i j} G_{i} \alpha_{j} \\
    &
    - \frac{1}{2} a^{3} \dot{\Phi}_{0} \dot{\Phi}_{0} \delta^{i j} \alpha_{i} \alpha_{j} 
    + \frac{1}{2} a^{3} \dot{\Phi}_{0} \dot{\Phi}_{0} \delta^{i j} G_{i} \alpha_{j} 
    + a^{3} V_{0} \delta^{i j} G_{i} \alpha_{j} 
    \\
    &
    - ( 2 c_{1} - c_{2} ) a^{3} \delta^{i j} \delta^{k l} \dot{C}_{i k} \dot{C}_{j l} 
    + 2 ( 2 c_{1} - c_{2} + 3 c_{3} ) H a^{3} \delta^{i j} \delta^{k l} C_{i k} \dot{C}_{j l} 
    + 2 ( 2 c_{1} - c_{2} ) a^{2} \delta^{i j} \delta^{k l} \partial_{i}{G_{k}} \dot{C}_{j l} \\
    &
    - ( 2 c_{1} - c_{2} ) a^{-1} \delta^{i j} \delta^{k l} \delta^{m n} \partial_{i}{C_{k m}} \partial_{n}{C_{j l}} 
    + ( 2 c_{1} - c_{2} ) a^{-1} \delta^{i j} \delta^{k l} \delta^{m n} \partial_{i}{C_{k m}} \partial_{j}{C_{l n}} 
    + c_{3} a^{-1} \delta^{i j} \delta^{k l} \delta^{m n} \partial_{i}{C_{j k}} \partial_{m}{C_{l n}} \\
    &
    - 2 c_{1} a^{-1} \delta^{i j} \delta^{k l} \partial_{i}{G_{k}} \partial_{j}{G_{l}} 
    + 4 c_{1} H^{2} a^{3} \delta^{i j} G_{i} G_{j} 
    \\
    &
    + \frac{1}{2} {a}^{3} \dot{\delta\Phi} \dot{\delta\Phi} 
    - \frac{1}{2}\ a \delta^{i j} \partial_{i}{\delta\Phi} \partial_{j}{\delta\Phi} 
    - \frac{1}{2} a^{3} V''_{0} \delta\Phi \delta\Phi 
    \,,
\end{split}
\label{VP of NGR with}
\end{equation}
Here, we have applied the background equations, Eqs.~\eqref{bcgd eqs of NGR with test matter field} and~\eqref{bcgd eq of test matter field}. 
We also set $\delta^{i j} \delta^{k l} \dot{C}_{i k} \dot{C}_{j l} = 2 \delta^{i j} \delta^{k l} \partial_{i}{\dot{C}_{k}} \partial_{j}{\dot{C}_{l}}$.

Varying with respect to $G_{i}$, we obtain a constraint. 
Substituting it back into Eq.~\eqref{VP of NGR with} and integrating out the decoupled term that does not contain any time derivative, we derive 
\begin{equation}
\begin{split}
    &\therefore\qquad 
    \mathcal{L}^\mathrm{(VP\,;\,2nd\,order)}_\mathrm{total} = \\
    &
    + ( 2 c_{1} - c_{2} + c_{3} ) a^{3} \delta^{i j} \dot{\alpha}_{i} \dot{\alpha}_{j} 
    + 2 ( 2 c_{1} - c_{2} + 3 c_{3} ) H a^{3} \delta^{i j} \alpha_{i} \dot{\alpha}_{j} 
    - 2 c_{1} a^{-1} \delta^{i j} \delta^{k l} \partial_{i}{\alpha_{k}} \partial_{j}{\alpha_{l}} 
    + 3 ( 2 c_{1} - c_{2} + 3 c_{3} ) H^{2} a^{3} \delta^{i j} \alpha_{i} \alpha_{j} 
    \\
    &
    + 2 c_{3} a^{2} \delta^{i j} \delta^{k l} \partial_{i}{C_{j k}} \dot{\alpha}_{l} 
    + 2 ( 2 c_{1} - c_{2} + 3 c_{3} ) H a^{2} \delta^{i j} \delta^{k l} \alpha_{i} \partial_{k}{C_{j l}} 
    - \frac{1}{2} a^{3} \dot{\Phi}_{0} \dot{\Phi}_{0} \delta^{i j} \alpha_{i} \alpha_{j} 
    \\
    &
    - ( 2 c_{1} - c_{2} ) a^{3} \delta^{i j} \delta^{k l} \dot{C}_{i k} \dot{C}_{j l}  
    + 2 ( 2 c_{1} - c_{2} + 3 c_{3} ) H a^{3} \delta^{i j} \delta^{k l} C_{i k} \dot{C}_{j l} 
    - ( 2 c_{1} - c_{2} ) a^{-1} \delta^{i j} \delta^{k l} \delta^{m n} \partial_{i}{C_{k m}} \partial_{n}{C_{j l}} 
    \\
    &
    + ( 2 c_{1} - c_{2} ) a^{-1} \delta^{i j} \delta^{k l} \delta^{m n} \partial_{i}{C_{k m}} \partial_{j}{C_{l n}} 
    + c_{3} a^{-1} \delta^{i j} \delta^{k l} \delta^{m n} \partial_{i}{C_{j k}} \partial_{m}{C_{l n}} \\
    &
    + \frac{1}{2} {a}^{3} \dot{\delta\Phi} \dot{\delta\Phi} 
    - \frac{1}{2}\ a \delta^{i j} \partial_{i}{\delta\Phi} \partial_{j}{\delta\Phi} 
    - \frac{1}{2} a^{3} V''_{0} \delta\Phi \delta\Phi 
    \,;
\end{split}
\label{VP of NGR with constraint Gi}
\end{equation}
In Type~6 (TEGR), repeating the same procedure with respect to $\alpha_{i}$, Eq.~\eqref{VP of NGR with constraint Gi} reduces to 
\begin{equation}
\begin{split}
    &\therefore\qquad 
    \mathcal{L}^\mathrm{(VP\,;\,2nd\,order)}_\mathrm{Type\,6} = \\
    &
    - ( 2 c_{1} - c_{2} ) a^{3} \delta^{i j} \delta^{k l} \dot{C}_{i k} \dot{C}_{j l} 
    + 2 ( 2 c_{1} - c_{2} + 3 c_{3} ) H a^{3} \delta^{i j} \delta^{k l} C_{i k} \dot{C}_{j l} 
    - ( 2 c_{1} - c_{2} ) a^{-1} \delta^{i j} \delta^{k l} \delta^{m n} \partial_{i}{C_{k m}} \partial_{n}{C_{j l}} 
    \\
    &
    + ( 2 c_{1} - c_{2} ) a^{-1} \delta^{i j} \delta^{k l} \delta^{m n} \partial_{i}{C_{k m}} \partial_{j}{C_{l n}} 
    + c_{3} a^{-1} \delta^{i j} \delta^{k l} \delta^{m n} \partial_{i}{C_{j k}} \partial_{m}{C_{l n}} 
    \\
    &
    + \frac{1}{2} {a}^{3} \dot{\delta\Phi} \dot{\delta\Phi} 
    - \frac{1}{2}\ a \delta^{i j} \partial_{i}{\delta\Phi} \partial_{j}{\delta\Phi} 
    - \frac{1}{2} a^{3} V''_{0} \delta\Phi \delta\Phi 
    \,;
\end{split}
\label{VP of Type 6 NGR with constraint wrt alpha_i}
\end{equation}
At first glance, this result seems to indicate a contradiction. 
However, since $C_{i j}$ is not a gauge-invariant variable, an appropriate choice of gauge can remove this apparent mode.
Type~3 contains only modes that stem from the violation of local Lorentz invariance particularly in the boost sector (see Sec.~\ref{04-02}); 
only $\alpha_{i}$ should propagate. 

If we choose Gauge choice I, we obtain~\cite{Tomonari:2025kgb} 
\begin{equation}
\begin{split}
    &\mathcal{L}^\mathrm{(VP\,;\,2nd\,order)}_\mathrm{Type\,3} 
    \\
    &= 
    ( 2 c_{1} - c_{2} + c_{3} ) a^{3} \delta^{ij} \dot{\alpha}_{i} \dot{\alpha}_{j} + 4 ( 2 c_{1} - c_{2} + 3 c_{3} ) a^{3} H \delta^{ij} \dot{\alpha}_{i} \alpha_{j} 
    - 2 c_{1} a \delta^{ij} \delta^{kl} \partial_{i}{\alpha_{k}} \partial_{j}{\alpha_{l}} 
    - 3 ( c_{2} - 3 c_{3} ) a^{3} H^{2} \delta^{ij} \alpha_{i} \alpha_{j}
    \\
    &\qquad
    + \frac{1}{2} a^{3} \dot{\delta\Phi} \dot{\delta\Phi} 
    - \frac{1}{2} a \delta^{ij} \partial_{i}{\delta\Phi} \partial_{j}{\delta\Phi} 
    - \frac{1}{2} V''_{0} a^{3} \delta\Phi \delta\Phi 
    \,.
\end{split}
\label{VPs of Type 3 with gauge choice I in gauge invariant variable}
\end{equation}
We note that this Lagrangian is already expressed in terms of gauge-invariant variables and corresponds to Case A in our classification. 

The ghost free condition is determined only by the kinetic terms since the determinant of the kinetic matrix of this type is
\begin{equation}
\begin{split}
    &\mathcal{K}_\mathrm{VP} 
    := 
    \mathrm{det}\left(\frac{\delta^{2}\mathcal{L}^\mathrm{(VP\,;\,2nd\,order)}_\mathrm{Type\,3}(t\,,\vec{z})}{\delta\dot{X}_{i}(t\,,\vec{x})\delta\dot{Y}_{j}(t\,,\vec{y})}\right) 
    =  
    \frac{1}{2} a^{12} ( 2 c_{1} - c_{2} + c_{3} )^{3} 
    \Delta_\mathrm{VP}(\vec{x}\,,\vec{y}\,,\vec{z})
    \,,
    \\
    &\Delta_\mathrm{VP}(\vec{x}\,,\vec{y}\,,\vec{z}) 
    = 
    [\delta^{(3)}(\vec{z} - \vec{x}) \delta^{(3)}(\vec{z} - \vec{y})]^{4}
    \,,
\end{split}
\label{det of kinetic matrix of VPs}
\end{equation}
where $X_{i} = \{\alpha_{i}(t\,,\vec{x})\,,\delta\Phi(t\,,\vec{x})\}$. 
This condition is the same as that in the scalar perturbation up to a factor that consists only of the scale factor and its time derivatives, which does not affect the existence of ghost modes. 
We remark that the kinetic matrix considered here coincides with the non-degenerate block of the kinetic matrix of the unreduced Lagrangian given in Appendix~\ref{app:01}.

\section{\label{05}Background-hierarchy bound in Type~3 around the flat-FLRW spacetime}

\subsection{\label{05-01} Two types of strong coupling in theories of gravity: EFT strong-coupling versus perturbatively/background-hierarchy bound} 

Any perturbation theory has a limited scale of validity in energy level. 
Once the system exceeds this energy scale, the perturbative approximation ceases to be reliable.
Strong coupling represents a typical example of such a limitation. 
To illustrate this perspective, we briefly review strong coupling in the Proca theory~\cite{Hell:2021oea,Hell:2021wzm}. 
In this context, although strong coupling itself is not the main concern, it is rather invoked to resolve the discontinuity that arises in the massless limit. 

The Proca theory is described by  
\begin{equation}
\begin{split}
    &S_\mathrm{Proca} 
    = 
    \int \left( -\frac{1}{4} F_{\mu \nu} F^{\mu \nu} + \frac{1}{2} m^{2} A_{\mu} A^{\mu} \right) dx^{4} \,,\\
    &F_{\mu \nu} = \partial_{\mu}{A_{\nu}} - \partial_{\nu}{A_{\mu}} \,,
\end{split}
\label{Proca theory}
\end{equation}
where $m$ denotes the mass of the vector field.
Varying the action with respect to $A_{0}$ yields a constraint equation.
Substituting this constraint back into Eq.~\eqref{Proca theory} and decomposing the spatial components $A_{i}$ into a transverse vector $\tilde{A}_{i}$ and longitudinal scalar $\chi$, we obtain 
\begin{equation}
\begin{split}
    &S_\mathrm{Proca} 
    = 
    - \frac{1}{2} 
    \int 
    \left[ 
    \tilde{\chi} (\square + m^{2}) \tilde{\chi}
    + 
    \tilde{A}_{i} ( \square + m^{2} ) \tilde{A}_{i}
    \right] dx^{4} 
    \,,\\
    &
    \tilde{\chi} 
    = 
    m\sqrt{\frac{-\Delta}{-\Delta + m^{2}}} \chi \,,
\end{split}
\end{equation}
The operator $- \Delta / ( - \Delta + m^{2} )$ acting on $\chi$ is understood in Fourier space.
Since the action integral should be dimensionless, at a characteristic length scale $L \sim k^{-1}$, the typical magnitudes of the fields can be estimated as
\begin{equation}
    \delta \tilde{A}_{i} \sim L^{-1} 
\label{magnitude of A before strong coupling}
\end{equation}
and 
\begin{equation}
    \delta\chi \sim (mL)^{-1}
    \,.
\label{magnitude of chi before strong coupling}
\end{equation}
In particular, in the massless limit $m \to 0$, the fluctuation $\delta \chi$ diverges.
This signals the well-known discontinuity of the Proca theory.

To remedy this pathology, we add a quartic self-interaction term \textit{by hand} into the action as
\begin{equation}
\begin{split}
    &S_\mathrm{modified\,Proca} 
    = 
    S_\mathrm{Proca} 
    + 
    S_\mathrm{self\text{-}interaction} \,,\\
    &S_\mathrm{self\text{-}interaction}
    =
    \int \frac{g^{2}}{4}(A_{\mu}A^{\mu})^{2} dx^{4} 
    \,.\\
\end{split}
\label{Proca theory with self-interaction}
\end{equation}
Expanding the interaction term in terms of the canonically normalized fields, we obtain
\begin{equation}
    S_\mathrm{self\text{-}interaction} 
    = 
    \int \Big[
    \underbrace{
    \frac{g^{2}}{4m^{4}} (\partial_{\mu}{\tilde{\chi}}\partial^{\mu}{\tilde{\chi}})^{2} 
    }_{\mathrm{dominant\,term\,of\,}\tilde{\chi}}
    + \cdots
    \quad
    \underbrace{
    -\frac{g^{2}}{m^{3}} \tilde{A}_{i} \partial^{i}{\tilde{\chi}} \partial_{\mu}{\tilde{\chi}} \partial^{\mu}{\tilde{\chi}} 
    }_{\mathrm{dominant\,term\,of\,}\tilde{A}_{i}} 
    + \cdots
    \Big] dx^{4} 
    \,,
\label{self-interaction terms of Proca theory}
\end{equation}
where only the dominant interaction terms are shown.
Strong coupling sets in at the scale where the interaction term becomes comparable to, and eventually dominates over, the kinetic term.
For the longitudinal and transverse modes, the corresponding strong coupling scales are given by
\begin{equation}
    L^{\chi}_\mathrm{str.} 
    = 
    \frac{\sqrt{g}}{m} 
    \,
\label{strong coupling scale of chi}
\end{equation}
and 
\begin{equation}
    L^{A}_\mathrm{str.} 
    = 
    \frac{g^{2/3}}{m} 
    \,.
\label{strong couping scale of A}
\end{equation}
Therefore, at length scales $L < L^{\chi}_\mathrm{str.}$, the longitudinal mode $\tilde{\chi}$ enters the strong coupling regime and ceases to propagate. 
At scales satisfying $L < L^{A}_\mathrm{str.} < L^{\chi}_\mathrm{str.}$, the transverse mode $\tilde{A}_{i}$ likewise becomes strongly coupled and no longer propagates.

Once the longitudinal mode enters its strong coupling scale, the magnitude given in Eq.~\eqref{magnitude of chi before strong coupling} is modified. 
That is, we expand $\chi$ in powers of the coupling constant $g^{2}$ as
\begin{equation}
    \chi = \chi^{(0)} + \chi^{(1)} + \cdots 
    \,.
\label{expansion of chi}
\end{equation}
In the strong coupling regime, the leading contribution is no longer $\chi^{(0)}$ but $\chi^{(1)}$.
Accordingly, the dominant term in the self-interaction reduces to
\begin{equation}
    \frac{g^{2}}{4}(\partial_{\mu}{\chi^{(1)}}\partial^{\mu}{\chi^{(1)}})^{2}
    \in S_\mathrm{self\text{-}interaction} 
    \,,
\label{dominant term of chi1 in self-interaction after strong coupling}
\end{equation}
The magnitude of $\chi^{(1)}$ is then estimated as
\begin{equation}
    \delta\chi^{(1)} \sim \frac{1}{\sqrt{g}} \,, 
\label{magnitude of chi1 after strong coupling}
\end{equation}
indicating that the discontinuity is removed at the cost of rendering the longitudinal mode non-propagating.

In the regime $L^{A}_\mathrm{str.} < L < L^{\chi}_\mathrm{str.}$, the transverse mode does not enter its strong coupling scale. 
Since the dominant contribution involving $\tilde{A}_{i}$ in the self-interaction scales as $g^{2} \tilde{A}^{(0)}_{i}\partial^{i}{\tilde{\chi}^{(1)}}\partial_{\mu}{\chi^{(1)}}\partial^{\mu}{\chi^{(1)}}$, comparing its magnitude with that of the kinetic term yields
\begin{equation}
    g^{2} \tilde{A}^{(0)}_{i}\partial^{i}{\tilde{\chi}^{(1)}}\partial_{\mu}{\chi^{(1)}}\partial^{\mu}{\chi^{(1)}} 
    \sim 
    \frac{g^{1/2}}{L^{4}} < \frac{1}{L^{4}}
    \sim 
    \tilde{A}^{0}_{i} ( \square + m^{2} ) \tilde{A}^{(0)}_{i}
    \,,
\label{magnitude of A after strong coupling}
\end{equation}
where $\tilde{A}_{i}$ is likewise expanded in powers of $g^{2}$ as $\tilde{A}_{i} = \tilde{A}^{(0)}_{i} + \tilde{A}^{(1)}_{i} + \cdots$, 
with $\tilde{A}^{(0)}_{i}$ denoting the leading-order term of which magnitude is given by Eq.~\eqref{magnitude of A before strong coupling}.
This implies that the transverse mode continues to propagate at all scales. 
This treatment is essentially the same as that is presented in Ref.~\cite{Vainshtein:1972sx}, which is known as Vainstein mechanism. 

We have reviewed a brief example of strong coupling in particle physics.
In the presence of gravity, however, the situation changes. 
The background spacetime evolves in time and is no longer the Minkowski spacetime assumed in particle physics.
This difference gives rise to another type of strong coupling.
Indeed, when estimating the magnitudes of the longitudinal and transverse modes in Eqs.~\eqref{magnitude of A before strong coupling}, \eqref{magnitude of chi before strong coupling}, and~\eqref{magnitude of chi1 after strong coupling}, 
we implicitly assumed a static background spacetime.
In cosmology, by contrast, we consider the flat FLRW spacetime, whose background dynamics is described by 
\begin{equation}
    \mathcal{L}_\mathrm{Background} 
    = 
    - 3 {M_{pl}}^{2} \left(2 c_{1} - c_{2} + 3 c_{3}\right) a^{3} H^{2} 
    +\frac{1}{2}a^{3}\dot{\Phi}_{0}\dot{\Phi}_{0} - a^{3}V_{0}
    \,
\label{Background Lagrangian}
\end{equation}
with matter fields. 
In the Minkowski limit, $\mathcal{L}_\mathrm{Background} = a^{3}\dot{\Phi}_{0}\dot{\Phi}_{0}/2 - a^{3}V_{0}$ due to $H = 0$. 
However, since our universe is expanding, $H \neq 0$.
This implies that in cosmology the estimation of the magnitude of each perturbation field depends on the energy scale associated with the background spacetime evolution. 
If the background evolution dominates over a given perturbation, that perturbation modes cease to propagate --- not through the conventional strong coupling mechanism familiar from particle physics --- but due to the dominance of the background itself. 
We therefore state that cosmological perturbation theory remains well-behaved only when such a hierarchy violation between the background and perturbations is absent. 
To distinguish this issue from the conventional strong coupling known in particle physics, we refer to as `\textit{perturbatively/background-hierarchy bound}' around a background spacetime as mentioned in Sec.~\ref{01}. 
Therefore, in theories of gravity, there are two types of strong coupling. 

In the following subsections, we investigate the presence of background-hierarchy bound around the flat FLRW background for each perturbation field in Type~3 of NGR: $h_{ij}$, $\gamma$, and $\alpha_{i}$. 
From now on, we explicitly represent the reduced Planck mass as $M_{pl}$. 

\subsection{\label{05-02}Background-hierarchy bound in tensor perturbation} 

The perturbed Lagrangian is given by~\cite{Tomonari:2025kgb}
\begin{equation}
\begin{split}
    & {M_{pl}}^{-2} \mathcal{L}^\mathrm{(TP\,;\,2nd\,order)}_\mathrm{total} 
    = 
     - a^{3}\,\left(2\,c_{1}- c_{2}\right)\,\delta^{ik}\,\delta^{jl}\,\dot{h}_{ij}\,\dot{h}_{kl} 
     + \cdots
    + \frac{1}{2} a^{3} \dot{\delta\Phi} \dot{\delta\Phi} - \frac{1}{2} a \delta^{i j} \partial_{i}{\delta\Phi} \partial_{j}{\delta\Phi} 
    - \frac{1}{2} a^{3} V''_{0} \delta\Phi \delta\Phi 
    \,,
\end{split}
\label{TPs of NGR in 2nd-order: only dominant}
\end{equation} 
where we have displayed only the dominant and relevant terms of $h_{ij}$ for evaluating the background-hierarchy bound. 
The full expression is given in Ref.~\cite{Tomonari:2025kgb}. 
Hereinafter, we use the relation of $\delta\dot{h}_{ij} \sim h_{ij}L^{-1}$ for estimating the time derivatives in the comparison, where $L$ is the inverse of wave length: $L \sim k^{-1}$. 
Note that this perturbed Lagrangian is already written solely in terms of gauge-invariant variables.

The signature of the determinant of the kinetic matrix depends on $-(2c_{1} - c_{2})$ and does not vanish. 
Thus, the ghost-free condition is given by\footnote{
In terms of $SO(1,3)$-irreducible decomposition of NGR~\cite{Hayashi:1979qx,Bohmer2011,Bohmer2012,Blixt:2019ene,Bahamonde:2024zkb}, the parameters $c_{1},c_{2},c_{3}$ corresponds to $c_\mathrm{ten},c_\mathrm{vec},c_\mathrm{axi}$ in the following manner:
\begin{equation}
    c_\mathrm{ten} = -\frac{4}{3}c_{1} + \frac{2}{3}c_{2}\,,\quad 
    c_\mathrm{vec} = -\frac{2}{3}c_{1} + \frac{1}{3}c_{2} - c_{3}\,,\quad
    c_\mathrm{axi} = 6c_{1} + 6c_{2}
    \,.
\label{correspondence of parameter space}
\end{equation}
Remark that the sign of $c_{1}$ and $c_{3}$ are opposite to those in Ref.~\cite{Bahamonde:2024zkb}. 
In Ref.~\cite{Bahamonde:2024zkb}, the ghost-free condition for Type~3 is given by either $c_\mathrm{vec} > 0 \quad \mathrm{and} \quad c_\mathrm{ten} > 0$ or $-c_\mathrm{vec} > c_\mathrm{ten} > 0$. 
Applying the Type~3 condition, Eq.~\eqref{Type 3 parameter condition}, $c_\mathrm{ten}$ and $c_\mathrm{vec}$ reduce to 
\begin{equation}
    c_\mathrm{ten} = \frac{4}{3}c_{2} 
    \,,\qquad
    c_\mathrm{vec} 
    = \frac{2}{3}c_{2} - c_{3} 
    = - \frac{4}{3}c_{2} + (2c_{2} - c_{3}) 
    =-c_\mathrm{ten} + (2c_{2} - c_{3}) 
    \,.
\end{equation}
In the current set-up, $c_\mathrm{ten} > 0$ since we have Eq.~\eqref{ghost-free region of Type 3}. 
Thus, we obtain $-c_\mathrm{vec} = c_\mathrm{ten} + (c_{3} - 2c_{2}) > c_\mathrm{ten} > 0$ by virtue of Eq.~\eqref{ghost-free region of Type 3}, and it coincides with the ghost-free condition $-c_\mathrm{vec} > c_\mathrm{ten} > 0$; 
the current set-up is consistent with that in the previous work~\cite{Bahamonde:2024zkb}. 
}
\begin{equation}
    2 c_{1} - c_{2} < 0\,,
\label{ghost-free condition of tensor mode}
\end{equation}
while $c_{3}$ remains arbitrary.
In Type~3, the parameter condition Eq.~\eqref{Type 3 parameter condition}, $2 c_{1} + c_{2} = 0$, implies
\begin{equation}
    c_{2} > 0 \,.
\label{condition of c2}
\end{equation} 
If the parameter condition $2 c_{1} - c_{2} + c_{3} = 0$ holds, the theory results in Type~6. 
Thus, for consistency, the parameter space of Type~3 should contain that of Type~6 (TEGR). 
In Type~6 (TEGR), since in our setup $c_{2} = 1/2 > 0$, the present parameter space is consistent. 
Conversely, combining Eq.~\eqref{condition of c2} with the Type~3 parameter condition Eq.~\eqref{Type 3 parameter condition}, we find that $c_{1}$ must be negative:
\begin{equation}
    c_{1} < 0 \,. 
\label{condition of c1}
\end{equation}
In Type~6, $c_{1} = - 1/4 < 0$; 
the present parameter space is consistent with our setup. 

Comparing the dominant term in Eq.~\eqref{TPs of NGR in 2nd-order: only dominant} with the background Lagrangian Eq.~\eqref{Background Lagrangian}, the background-hierarchy bound is estimated as
\begin{equation}
    \delta h_{ij} 
    \sim 
    \sqrt{-\frac{2 c_{1} - c_{2} + 3 c_{3}}{2 c_{1} - c_{2}}} 
    = 
    \sqrt{- 1 + \frac{3 c_{3}}{2 c_{2}}} 
    \,,
\label{background strong coupling scale of tensor mode}
\end{equation}
where we have used Eq.~\eqref{Type 3 parameter condition}.
If the Type~8 limit, \textit{i.e.}, $c_{3} \rightarrow 2c_{2}/3$, could be taken the tensor modes was free from background-hierarchy bound around the flat FLRW spacetime. However, it is not the case since this limit violates the ghost-free condition of Type~3,
\begin{equation}
    c_{2} > 0\,,
    \quad
    c_{3} > 2 c_{2}
    \,;
\label{ghost-free region of Type 3}
\end{equation}
the tensor modes in Type~3 suffer from the background-hierarchy bound.

For cosmological applications, nevertheless, the perturbation theory of a given theory of gravity is generally viable as long as the background-hierarchy bound does not arise in the regime where inflation provides the initial conditions for the subsequent evolution of the universe.
Therefore, if there exists a non-empty parameter region $(c_{2}\,,c_{3})$ such that
\begin{equation}
    0 < c_{3} - \frac{2}{3} c_{2}  < \epsilon \,
\label{condition for free from strong coupling}
\end{equation}
for a positive valued parameter $\epsilon$, tensor perturbations can consistently describe our universe. 
As far as $\epsilon>0$, there always exists a parameter space $(c_{2},c_{3})$ consistent with the ghost-free region of Type~3: Eq.~\eqref{ghost-free region of Type 3}. 

For $\delta h_{ij} \sim (L M_{pl})^{-1} \lesssim \sqrt{\epsilon}$, taking the upper bound of Eq.~\eqref{condition for free from strong coupling}, we estimate $c_{2}$ and $c_{3}$ as 
\begin{equation}
    c_{2} 
    = 
    \frac{3}{2}L^{2}M_{pl}^{2} \epsilon 
    \,,\qquad
    c_{3} 
    = 
    \left( 1 + L^{2} M_{pl}^{2} \right) \epsilon 
    \,. 
\label{condition of c2 and c3 when delta h = planck mass}
\end{equation}

\subsection{\label{05-03}Background-hierarchy bound in vector perturbation} 

The perturbed Lagrangian is given by~\cite{Tomonari:2025kgb}
\begin{equation}
\begin{split}
    &{M_{pl}}^{-2} \mathcal{L}^\mathrm{(VP\,;\,2nd\,order)}_\mathrm{total} 
    = 
    ( 2 c_{1} - c_{2} + c_{3} ) a^{3} \delta^{i j} \dot{\alpha}_{i} \dot{\alpha}_{j} 
    + \cdots
    + \frac{1}{2} a^{3} \dot{\delta\Phi} \dot{\delta\Phi} 
    - \frac{1}{2} a \delta^{ij} \partial_{i}{\delta\Phi} \partial_{j}{\delta\Phi} 
    - \frac{1}{2} V''_{0} a^{3} \delta\Phi \delta\Phi 
    \,;
\end{split}
\label{further simplified VPs and pseudo-VPs of NGR with time-developing matter in Type 3 and Type 7}
\end{equation}
where we have retained only the dominant and relevant terms of $\alpha_{i}$ associated with background-hierarchy bound. 
The full expression is given in Ref.~\cite{Tomonari:2025kgb}. 
This Lagrangian is written solely in terms of gauge-invariant variables. 
Here, we note that the kinetic term of the reduced Lagrangian, obtained by substituting the constraint into the original Lagrangian, is identical to that of the original Lagrangian in the present theory. 
Therefore, the kinetic term in the reduced Lagrangian can be used to estimate the background-hierarchy bound, since the bound is relevant only to the kinetic term of the theory in the present theory. 
Explicitly, compare the kinetic term of the unreduced Lagrangian in Appendix~\ref{app:01} with that in Sec.~\ref{04}.

The ghost-free condition is given by 
\begin{equation}
    2 c_{1} - c_{2} + c_{3} = - 2c_{2} + c_{3} > 0 \,,
\label{ghost-free condition of lorentz vector}
\end{equation} 
which coincides with that for the scalar perturbation, as shown in the next subsection and also investigated in our previous work~\cite{Tomonari:2025kgb}, 
where we have used Eq.~\eqref{Type 3 parameter condition}. 

The background-hierarchy bound is estimated as 
\begin{equation}
    \delta \alpha_{i} 
    \sim \sqrt{\frac{2 c_{1} - c_{2} + 3 c_{3}}{2 c_{1} - c_{2} + c_{3}}} 
    = 
    \sqrt{1 - \frac{2 c_{3}}{2 c_{2} - c_{3}}} 
    \,,
\label{background strong coupling scale of vector mode}
\end{equation}
where we have used Eq.~\eqref{Type 3 parameter condition}.
If there exists a parameter $\epsilon$ such that Eq.~\eqref{condition for free from strong coupling} is satisfied, the vector mode remains free from background-hierarchy bound around the flat FLRW spacetime.

\subsection{\label{05-04}Background-hierarchy bound in scalar perturbation} 

The perturbed Lagrangian is given by Eq.~\eqref{SPs in Type 3 with gauge choice I in gauge invariant variable} as 
\begin{equation}
\begin{split}
    &{M_{pl}}^{-2} \mathcal{L}^\mathrm{(SP\,;\,2nd\,order)}_\mathrm{Type\,3} 
    = 
    ( 2 c_{1} - c_{2} + c_{3} ) a H^{-2} \delta^{i j} \partial_{i}{\dot{\gamma}} \partial_{j}{\dot{\gamma}}  
    + \cdots 
    + \frac{1}{2} a^{3} \dot{\delta\Phi} \dot{\delta\Phi} 
    - \frac{1}{2} a \delta^{ij} \partial_{i}{\delta\Phi} \partial_{j}{\delta\Phi} 
    - \frac{1}{2} a^{3} V''_{0} \delta\Phi \delta\Phi 
    \,,
\end{split}
\label{SPs in Type 3 with gauge choice I in gauge invariant variable: only dominant}
\end{equation}
where we have retained only the dominant and relevant terms of $\gamma$ associated with the background-hierarchy bound in this sector. 
The full expression is given in Ref.~\cite{Tomonari:2025kgb}. 
The kinetic term is, of course, identical to that in the unreduced Lagrangian given in Ref.~\cite{Tomonari:2025kgb} after translating $\alpha$ by $\gamma$ with Eq.~\eqref{scalar gauge invariant variables in Gauge choice I}. 

The ghost-free condition is
\begin{equation}
    2 c_{1} - c_{2} + c_{3} > 0 \,,
\label{ghost-free condition of lorentz scalar}
\end{equation}
which coincides with that for the vector perturbation, as shown in the previous subsection and also investigated in our previous work~\cite{Tomonari:2025kgb}. 
Due to Eq.~\eqref{ghost-free condition of tensor mode}, $c_{3}$ must be positive:
\begin{equation}
    c_{3} > 0 \,,
\label{condition of c3}
\end{equation}
and this condition is consistent with Eq.~\eqref{ghost-free region of Type 3}. 
In Type~6, $c_{3} = 1 > 0$; 
the present parameter space is also consistent with our setup. 

The background-hierarchy bound is estimated as
\begin{equation}
    \delta \gamma 
    \sim 
    \sqrt{\frac{2 c_{1} - c_{2} + 3 c_{3}}{2 c_{1} - c_{2} + c_{3}}} 
    = 
    \sqrt{1 - \frac{2 c_{3}}{ 2 c_{2} - c_{3}}} 
    \,,
\label{background strong coupling scale of scalar mode}
\end{equation}
where we have used Eq.~\eqref{Type 3 parameter condition}.
This scale coincides with that of the vector modes.

For cosmological applications, the background-hierarchy bound of the scalar and vector modes must be larger than that of the tensor modes.
Otherwise, this perturbed theory is no longer a theory of gravity. 
The required condition is
\begin{equation}
\begin{split}
    &\delta h_{ij} = \sqrt{- 1 + \frac{3 c_{3}}{2 c_{2}}} 
    < 
    \sqrt{1 - \frac{2 c_{3}}{ 2 c_{2} - c_{3}}} = \delta \alpha_{i} = \delta\gamma
    \,,
\end{split}
\label{condition for the existence of tensor modes}
\end{equation}
or equivalently, 
\begin{equation}
    2c_{2} < c_{3} < 4 c_{2}
    \,.
\label{condition for the existence of tensor modes}
\end{equation}
Notice that this condition contains the ghost-free condition of Eq.~\eqref{ghost-free region of Type 3}. 
Therefore, the linear perturbation theory of Type~3 of NGR is viable as long as a non-empty overlap region of Eqs.~\eqref{ghost-free region of Type 3}, \eqref{condition for free from strong coupling}, and~\eqref{condition for the existence of tensor modes} exists. 
This existence is obvious for a positive valued parameter $\epsilon$. 
However, the Type~8 limit cannot be taken.  

\begin{figure}
    \centering
    \includegraphics[width=0.8\linewidth]{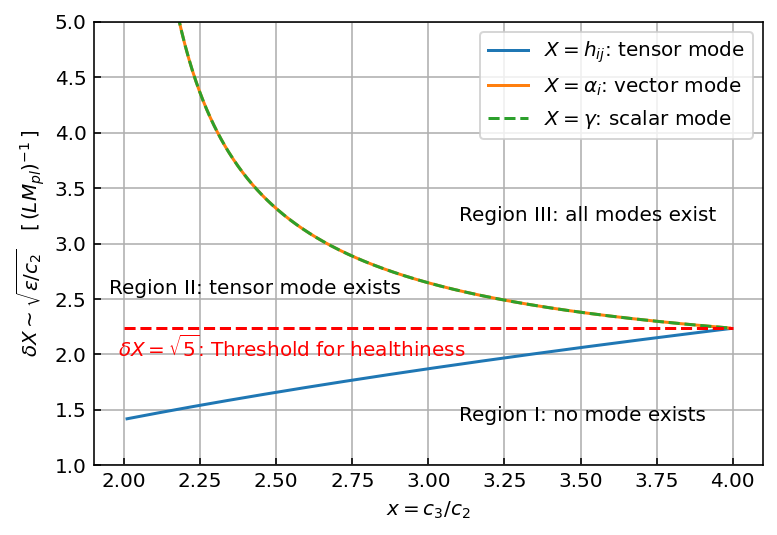}
    \caption{\textbf{Background-hierarchy bounds in Type~3 of NGR are shown in a single figure.}\\
    All the bounds intersect at $(x,\delta X)=(4,\sqrt{5})$. 
    The scalar and vector bounds diverge at $x=2$. 
    These bounds divide the required region $2<x<4$ into three distinct regions. 
    In Region~I, all propagating modes are subject to the background-spacetime development, indicating that no linearly viable perturbed model of Type~3 NGR exists in this region. 
    In Region~II, the tensor mode remains healthily behaved, whereas the vector and scalar modes are still bounded by the background-spacetime development. 
    As a remark on Regions~I and II, if a conventional EFT strong coupling can modify the relevant fundamental scale and remedy this pathology, the corresponding perturbed model of Type~3 NGR is applicable to cosmology, although it would be only within a strictly restricted regime. 
    In Region~III, all modes remain healthily behaved; therefore, perturbed models of Type~3 NGR are theoretically viable for cosmological applications at linear level.
    }
    \label{fig:BHBinType3}
\end{figure}

Finally, we describe all the background-hierarchy bounds in a single figure to clarify valid perturbed models. 
As stated in Ref.~\cite{Blixt:2018znp}, normalizing the three-free parameters, each Type of NGR is restricted by the parameters that are reduced its number by one. 
In the current situation, we can see this reduction by introducing a variable
\begin{equation}
    x = \frac{c_{3}}{c_{2}}
    \,.
\label{ratio of parameters in Type 3}
\end{equation}
Using this variable, the ghost-free condition, Eq.~\eqref{ghost-free region of Type 3}, and the required condition, Eq.~\eqref{condition for the existence of tensor modes}, are expressed in a single inequality
\begin{equation}
    2 < x < 4
    \,.
\label{inequality of ghost-free region and required condition}
\end{equation}
For the region $x<2$, models still stand at Type~3 but suffer from ghost instability.
For the region $4<x$, the ghost instability is always absent. 
However, there occurs a case in which the tensor mode does not propagate due to the background hierarchy bound. 
In this case, we do not consider Type~3 NGR as a theory of gravity. 
Thus, we cut off this region from our consideration. 
With this domain, we describe all the background-hierarchy bounds by the single parameter $x$ as follows: 
\begin{equation}
    \delta h_{ij} 
    \sim 
    \sqrt{- 1 + \frac{3}{2}x} 
    \,,\qquad
    \delta \alpha_{i} = \delta\gamma
    \sim 
    \sqrt{1 - \frac{2 x}{2 - x}}
    \,.
\label{delta h_ij, delta alpha_i, delta gamma in terms of x}
\end{equation}
We show all the bounds in a single Fig.~\ref{fig:BHBinType3}. 
Taking Eq.~\eqref{condition for free from strong coupling} into account, we obtain the lower bound of the parameter $\epsilon / c_{2}$: 
\begin{equation}
    \frac{4}{3} < \frac{\epsilon}{c_{2}}
    \,.
\label{condition for free from strong coupling in terms of x}
\end{equation}
Moreover, there is a threshold for the healthiness of the perturbed modes;
if $5 < \epsilon/c_{2}$, there is a perturbed model which is free from the background-hierarchy bounds. 
In contrast, if $5 > \epsilon/c_{2}$, all perturbed modes suffer from the background-hierarchy bounds at least for the scalar and vector mode. 
The magnitude of this parameter $x=c_{3}/c_{2}$ can be constrained in the relationship to physical variables by requiring consistency between the theoretical predictions and observational data, thereby providing an estimation of the valid domain of perturbed Type~3 NGR models. 

\paragraph{Remark.}
If Eq.~\eqref{SPs in Type 3 with gauge choice I in gauge invariant variable: only dominant} were not written in terms of gauge-invariant variables, that is, if one instead adhered to Eq.~\eqref{SPs in Type 3 with gauge choice I}, one would obtain an entirely incorrect background-hierarchy bound: 
\begin{equation}
    \delta \alpha 
    \sim L \sqrt{\frac{2 c_{1} - c_{2} + 3 c_{3}}{2 c_{1} - c_{2} + c_{3}}} 
    \sim \frac{1}{k} \sqrt{1 + \frac{2 c_{3}}{2 c_{1} - c_{2} + c_{3}}} 
    \rightarrow 0 \qquad (k\rightarrow\infty: \mathrm{UV-limit}) 
    \,;
\label{background strong coupling scale of scalar mode}
\end{equation}
which would incorrectly suggest the absence of background-hierarchy bound.
Thus, while Eq.~\eqref{SPs in Type 3 with gauge choice I} can be used to count the number of propagating degrees of freedom, it is no longer reliable for estimating the background-hierarchy bound. 
We expect the the same statement holds for the conventional EFT strong couplings.

From this observation, we finally emphasize the following:
\begin{quote}
    \textit{In theories of gravity, a strong coupling scale is sometimes identified incorrectly when the perturbed Lagrangian is not written in a gauge-invariant form}.
\end{quote}
We conclude that, as long as diffeomorphism invariance is preserved, Gauge choices I and II listed in Table~\ref{Table:PropagatingModesImPreferableGaugeChoice} of Sec.~\ref{03-03} can be consistently applied both to counting propagating degrees of freedom and to surveying background-hierarchy bounds around the flat FLRW spacetime.

\section{\label{06} Conclusions}

In this paper, we have investigated background-hierarchy bounds in Type~3 of NGR around the flat FLRW spacetime by revisiting the propagating modes with a preferable gauge choice. 
First, we examined how symmetry-breaking gives rise to a DOF in the context of DB analysis.
We then identified two fundamental patterns for generating a DOF. 
Based on this classification, we showed that there are two preferable gauge choices in a linear perturbation theory. 
The results are summarized in Table~\ref{Table:PropagatingModesImPreferableGaugeChoice}. 
For each gauge choice, we investigated the explicit form of gauge-invariant variables. 
We found that, when a preferable gauge choice is adopted, every perturbed theory can be written solely in terms of gauge-invariant variables, and that if at least boost invariance is violated, an Ostrogradsky instability generically arises in Case B of Gauge choice I. 
In particular, Case A in Gauge choice I allows us to capture a propagating scalar mode even if the perturbed Lagrangian is not written solely in terms of gauge-invariant variables, and in principle, there is no Ostrogradsky instability. 
Second, we revisited the linear perturbation theory of Type~3 NGR using an \textit{un-}preferable gauge choice, in order to confirm that the previous analysis in Ref.~\cite{Tomonari:2025kgb} correctly captures the number of propagating modes, in which Gauge choice I is adopted.
We also verified the absence of ghost modes in this type. 
Finally, after clarifying the distinctive features of strong coupling in gravitational theories, we estimated the background-hierarchy bound of the scalar, vector, and tensor modes in Type~3 of NGR.
We showed that there exists a non-empty parameter space in which the perturbed theory remains viable for cosmological applications, even though a background-hierarchy bound appears in the linear perturbation theory of Type~3 NGR.

In the previous works, perturbative analyses of both $f(T)$ gravity~\cite{Chen:2010va,Dent:2010nbw,Wu:2012hs,Izumi:2012qj} and NGR~\cite{Golovnev:2023ddv,Golovnev:2023jnc,Bahamonde:2024zkb,Tomonari:2025kgb}, including our own work~\cite{Tomonari:2025kgb}, have not always been carried out fully in terms of gauge-invariant variables. 
Moreover, in some of the existing literature, the implications of the emergence of propagating modes from a violation of local Lorentz symmetry in DB analysis were not discussed in detail. 
The linear perturbation theory of MAG with Weitzenb\"{o}ck gauge has been developed in the context of $f(T)$-gravity. 
In Refs.~\cite{Chen:2010va,Dent:2010nbw,Wu:2012hs}, the antisymmetric components of vierbein field associated with local Lorentz transformations were incorporated into the perturbative framework.  
In Ref.~\cite{Izumi:2012qj}, the antisymmetric tensor perturbations were further decomposed into a pseudo-scalar mode and transverse vector modes, and the gauge transformation laws of the perturbation fields were derived. 
However, the gauge conditions adopted in these analyses did not properly take into account the violation of local Lorentz invariance in $f(T)$-gravity. 
Moreover, the perturbative analyses were not performed fully in terms of gauge-invariant variables. 

In the context of NGR, we encountered a situation similar to that in $f(T)$ gravity.
In Ref.~\cite{Golovnev:2018wbh}, the linear perturbation theory of MAG in the Weitzenb\"{o}ck gauge was independently established, where the local Lorentz symmetry had already been incorporated into the theory. 
Subsequently, in Refs.~\cite{Golovnev:2023ddv,Golovnev:2023jnc}, perturbative analyses were performed around both Minkowski and flat FLRW spacetimes.
However, since these analyses neither took the results of the DB analysis of NGR into account nor were carried out in a gauge-invariant manner, their results do not necessarily respect the symmetry in this context. 
In our previous work~\cite{Tomonari:2025kgb}, we performed a perturbative analysis based on the results of the DB analysis~\cite{Blixt:2018znp,Tomonari:2024ybs,Tomonari:2024lpv}. 
Depending on the pattern of symmetry-breaking in NGR, we adopted appropriate gauge conditions and counted the number of propagating modes for each type of NGR. 
However, this analysis was not performed in a gauge-invariant manner, raising the concern as to whether our perturbative treatment correctly counted the number of propagating modes of the theory. 
In the present work, we have resolved this remaining issue by establishing a systematic procedure to rewrite the perturbed theory exclusively in terms of gauge-invariant variables, as summarized in Table~\ref{Table:PropagatingModesImPreferableGaugeChoice}. 
This enables us to confirm that our previous results in Ref.~\cite{Tomonari:2025kgb} remain unchanged regarding the counting of propagating modes. 
By contrast, we have also clarified that in theories of gravity, when estimating the background-hierarchy bound against a given background spacetime, it is essential to express the perturbed Lagrangian solely in terms of gauge-invariant variables. 
Otherwise, one may obtain a qualitatively incorrect background-hierarchy bound, as stated at the end of Sec.~\ref{05-04}.

For future work, introducing source terms and establishing a set of Boltzmann equations to describe the structure formation of the universe, we should investigate the phenomenological consistency of the linear perturbation theory of Type~3 NGR. 
We expect that this theory would provide a new perspective on cosmological issues such as $H_{0}$- and/or $S_{8}$-tension, dark matter, and dark energy.
In particular, we should focus our analysis on observationally restricting the magnitude of the parameter $c_{3}/c_{2}$ by relating to physical parameters in cosmological observations to prevent the background-hierarchy bounds in this perturbation theory. 
This would be achieved by modifying the methodology developed in Ref.~\cite{Hogas:2021lns}. 
If this region exists such that all the background-hierarchy bounds are smaller than that of the least scale that this perturbation theory should work in a healthy manner, this theory is applicable to cosmology. 
(See Sec.~\ref{05}.) 
Once this point is clarified, including higher-order perturbations, we should study the conventional EFT strong-coupling regime of each mode. 
All the ingredients for constructing EFT models of Type~3 NGR are now in place --- namely, the field contents: $h_{ij}$, $\alpha_i$, and $\gamma$, the symmetry: $SO(3)$, and the expansion parameters: one of $c_1$, $c_2$, $c_3$ with the valid region of these parameters (see Fig.~\ref{fig:BHBinType3}) and the coupling constants for higher-order EFT terms --- we are ready to proceed. 
In this future investigation, from the point of view of consistency relating to our forthcoming study we should pay attention to the pioneering works~\cite{Mikura:2021clt,He:2022xef,Mikura:2024fgp} in the context of Higgs inflation. 
We should also remark on the previous study~\cite{BeltranJimenez:2019nns}. 
In this work, revisiting the perturbations of NGR up to third order, it is stated that NGR can be consistent with a stable Minkowski background. 
We should revisit this statement in this future investigation. 

It would also be worthwhile to revisit the linear perturbation analysis of $f(T)$ gravity using the gauge choices summarized in Table~\ref{Table:PropagatingModesImPreferableGaugeChoice}, particularly with regard to the background-hierarchy bound around the flat FLRW background.
In the context of Refs.~\cite{Mikura:2021clt,He:2022xef,Mikura:2024fgp}, it would be interesting to explore a hybrid Type~9 scenario combined with Higgs inflation, in which tensor, (Higgs) scalar, and pseudo-scalar and pseudo-vector modes associated with Type~9 would coexist at the linear perturbative level. 
We leave these issues for future work.

\acknowledgments
K.T. would like to thank Shin'ichi Nojiri and Taishi Katsuragawa for beneficial discussions. S.B. and D.B. would like to thank Konstantinos F. Dialektopoulos and Anamaria Hell for useful discussions.
K.T. thanks Yunwei Yu for supporting my research activities. 
D.B. is supported by Carl Tryggers foundation, CTS 24:3698. 
S.B. is supported by the Institute for Basic Science under the project code IBS-R018-D3. 
We utilized Cadabra~\cite{Peeters:2007wn} for completing all calculations of this work. 
Finally, we would like to especially thank Alexey Golovnev for commenting on our work as the celebratory note in arXiv~\cite{Golovnev:2026xjc}. 

\appendix
\section{\label{app:01} Vector modes in Type~3 around FLRW background spacetime}

In Ref.~\cite{Golovnev:2026xjc}, the author claimed that the vector modes are spurious modes. 
That is, the reduced Lagrangian, namely the Lagrangian obtained after substituting a constraint, does not generically describe the true nature of the theory. 
This statement itself is correct. 
However, when we restrict ourselves to merely counting the number of propagating modes, the reduced Lagrangian remains a valid tool. 
In fact, the reduced Lagrangian is utilized in the literature, for instance, see Refs.~\cite{Bahamonde:2024zkb,Li:2020xjt,Li:2021wij,Hell:2021wzm,Hell:2021oea,Li:2022mti,Hell:2023rbf,Hell:2023mph,Hell:2024xbv,Kang:2025rxn,Chen:2025mhu,DeFelice:2025ykh,Hell:2026blj,Gotzberger:2026ujl}. 
In Ref.~\cite{Tomonari:2025kgb}, as stated in Introduction of the paper, they also aimed only to count the number of propagating modes in advance. 
As also mentioned in Conclusion of the paper, investigating the dynamics of each propagating mode is left for future work. 
In this investigation, of course, one should start with the unreduced Lagrangian. 
The key point is that, when one aims to go beyond just counting the number of propagating modes and clarifies the dynamics of these modes, one should not confuse the unreduced Lagrangian with its reduced counterpart. 
Instead, one must analyze the equations of motion derived from the unreduced Lagrangian. 

We verify these statements explicitly. 
The unreduced Lagrangian of Type~3 is given as follows: 
\begin{equation}
\begin{split}
    &\mathcal{L}^\mathrm{(VP;Type\,3)}_\mathrm{total} 
    \\
    &= 
    ( - 2 c_{2} + c_{3} ) a^{3} \delta^{ij} \dot{\alpha}_{i} \dot{\alpha}_{j} 
    - c_{2} a \delta^{kl} \alpha_{k} \Delta{\alpha_{l}} 
    \\
    &\qquad
    \underset{\mathrm{FLRW\,contribution}}{\underbrace{
    + 4 ( - 2 c_{2} + 3 c_{3} ) a^{3} H \delta^{ij} \dot{\alpha}_{i} \alpha_{j} 
    - 3 ( c_{2} - 3 c_{3} ) a^{3} H^{2} \delta^{ij} \alpha_{i} \alpha_{j} 
    }}
    \quad
    \underset{\mathrm{matter\,contribution}}{\underbrace{
    - \frac{1}{2} a^{3} \delta^{ij} \alpha_{i} \alpha_{j} \dot{\Phi}_{0} \dot{\Phi}_{0}
    }} 
    \\
    &\qquad 
    + G_{i} \Bigg{[} 
    + 2 c_{2} a \delta^{ij} \Delta{\alpha_{j}} 
    - c_{2} a \delta^{ij} \Delta{G_{j}} 
    \\
    &\qquad \qquad \quad 
    \underset{\mathrm{FLRW\,contribution}}{\underbrace{
    - 2 ( - 2 c_{2} + 3 c_{3} ) a^{3} H \delta^{ij} \dot{\alpha}_{j} 
    + 2 ( - 2 c_{2} + 3 c_{3} ) a^{2} H \epsilon^{ijk} \partial_{j}{\tilde{V}_{k}} 
    + 12 c_{2} a^{3} H^{2} \delta^{ij} G_{j} 
    - 12 c_{3} a^{3} H^{2} \delta^{ij} \alpha_{j} 
    }}
    \\
    &\qquad \qquad \quad 
    \underset{\mathrm{matter\,contribution}}{\underbrace{
    + a^{3} V_{0} \delta^{ij} \alpha_{j} 
    + \frac{1}{2} a^{3} \delta^{ij} \alpha_{j} \dot{\Phi}_{0} \dot{\Phi}_{0}
    }}
    \Bigg{]} 
    \\
    &\qquad 
    + \tilde{V}_{i} \Bigg{[}
    - ( - 2 c_{2} + c_{3} ) a \delta^{ij} \Delta{\tilde{V}_{j}} 
    + 2 ( 2 c_{2} - c_{3} ) a^{2} \epsilon^{ijk} \partial_{j}{\dot{\alpha}_{k}} 
    \\
    &\qquad \qquad \quad
    \underset{\mathrm{FLRW\,contribution}}{\underbrace{
    - 4 ( - 2 c_{2} + 3 c_{3} ) a^{3} H \delta^{ij} \dot{\tilde{V}}_{j} 
    + 4 ( - 2 c_{2} + 3 c_{3} ) a^{2} H \epsilon^{ijk} \partial_{k}\alpha_{j} 
    - 3 ( - 2 c_{2} + 3 c_{3} ) a^{3} H^{2} \delta^{ij} \tilde{V}_{j}
    }}
    \\
    &\qquad \qquad \quad
    \underset{\mathrm{matter\,contribution}}{\underbrace{
    - a^{3} V_{0} \delta^{ij} \tilde{V}_{j} 
    + \frac{1}{2}a^{3}\delta^{ij}\tilde{V}_{j}\dot{\Phi}_{0}\dot{\Phi}_{0}
    }}
    \Bigg{]}
    \\
    &\qquad 
    + \frac{1}{2} a^{3} \dot{\delta\Phi} \dot{\delta\Phi} 
    - \frac{1}{2} a \delta^{ij} \partial_{i}{\delta\Phi} \partial_{j}{\delta\Phi} 
    - \frac{1}{2} V''_{0} a^{3} \delta\Phi \delta\Phi 
    \,,
\end{split}
\label{unreduced Lagrangian of (pseudo-)vector mode in Type 3}
\end{equation}
where we have adopted the Type~3 condition, $2c_{1}+c_{2}=0$, integrated by parts and neglected total divergent terms; 
so far we have not performed any other manipulation including the substitution of a constraint. 
The kinetic matrix with respect to $\alpha_{i}$, $G_{i}$, and $\tilde{V}_{i}$ is given as follows:
\begin{equation}
    \mathcal{K}_{ij}
    =
    \frac{\delta^{2}\mathcal{L}^\mathrm{(VP;Type\,3)}_\mathrm{total}(t,\vec{z})}{\delta\dot{X}_{i}(t,\vec{x})\delta\dot{X}_{j}(t,\vec{y})}
    =
    \begin{bmatrix}
    (-2c_{2}+c_{3})a^{3}\mathbf{1}_{3\times3} & \mathbf{0}_{3\times3} & \mathbf{0}_{3\times3} & 0
    \\
    \mathbf{0}_{3\times3} & \mathbf{0}_{3\times3} & \mathbf{0}_{3\times3} & 0
    \\
    \mathbf{0}_{3\times3} & \mathbf{0}_{3\times3} & \mathbf{0}_{3\times3} & 0
    \\
    0 & 0 & 0 & \frac{1}{2}a^{3}
    \end{bmatrix}
    \delta^{(3)}(\vec{z}-\vec{x})\delta^{(3)}(\vec{z}-\vec{y})
    \,,
\label{kinetic matrix in Type 3}
\end{equation}
where $X_{i}\in\{\alpha_{i},G_{i},\tilde{V}_{i},\delta\Phi\}$. 
We have denoted $\mathbf{1}_{3\times3}=\delta_{ij}$ and $\mathbf{0}_{3\times3}=0\times\delta_{ij}$. 
Thus, when excepting for the scalar matter sector, there are six constraints and three solutions possible to propagate in Type~3 regardless of vacuum Minkowski or FLRW background spacetime. 
The latter implication is also obtained from the reduced Lagrangian. 
In the reduced Lagrangian, the kinetic matrix is given as a non-degenerate one in the present theory. (See Secs.~\ref{04-02} and~\ref{04-03}.) 
This indicates that the background-hierarchy bound is valid even under the criticism~\cite{Golovnev:2026xjc}.

To distinguish these three solutions from an undesirable one for cosmological applications, we need to investigate the equations of motion of Type~3. 
That is, we are of interest only in an oscillatory solution. 
Varying with respect to $\alpha_{i}$, $G_{i}$, and $\tilde{V}_{i}$, we obtain the equations of motion as follows: 
\begin{equation}
\begin{split}
    &
    -2(-2c_{2}+c_{3})a^{3}\delta^{ij}\ddot{\alpha}_{j} 
    -2c_{2}a\delta^{ij}\Delta\alpha_{j} 
    +2c_{2}a\delta^{ij}\Delta G_{j} 
    +2(2c_{2}-c_{3})a^{2}\epsilon^{jki}\partial_{k}\dot{\tilde{V}}_{j} 
    =
    \\
    &\qquad
    \underset{\mathrm{FLRW\,contribution}}{\underbrace{
    +6(-2c_{2}+c_{3})Ha^{3}\delta^{ij}\dot{\alpha}_{j} 
    +6(-3c_{2}+3 c_{3})a^{3}H^{2}\delta^{ij}\alpha_{j} 
    -2(-2c_{2}+3c_{3})a^{3}H\delta^{ij}\dot{G}_{j} 
    -6(-2c_{2}+c_{3})a^{3}H^{2}\delta^{ij}G_{j} 
    }}
    \\
    &\qquad
    \underset{\mathrm{matter\,contribution}}{\underbrace{
    +a^{3}\delta^{ij}\alpha_{j}\dot{\Phi}_{0}\dot{\Phi}_{0} 
    +4(-2c_{2}+3c_{3})a^{3}\dot{H}\delta^{ij}\alpha_{j} 
    -a^{3}V_{0}\delta^{ij}G_{j} 
    -\frac{1}{2}a^{3}\delta^{ij}G_{j}\dot{\Phi}_{0}\dot{\Phi}_{0} 
    -2(-2c_{2}+3c_{3})a^{3}\dot{H}\delta^{ij}G_{j} 
    }}
    \,,
    \\\\
    &
    2c_{2}a\delta^{ij}\Delta{\alpha_{j}} 
    -2c_{2}a\delta^{ij}\Delta{G_{j}} 
    = 
    \\
    &\qquad
    \underset{\mathrm{FLRW\,contribution}}{\underbrace{
    2(-2c_{2}+3c_{3})a^{3}H\delta^{ij}\dot{\alpha}_{j} 
    +12c_{3}a^{3}H^{2}\delta^{ij}\alpha_{j} 
    -24c_{2}a^{3}H^{2}\delta^{ij}G_{j} 
    -2(-2c_{2}+3c_{3})a^{2}H\epsilon^{ijk}\partial_{j}{\tilde{V}_{k}} 
    }}
    \\
    &\qquad
    \underset{\mathrm{matter\,contribution}}{\underbrace{
    -a^{3}V_{0}\delta^{ij}\alpha_{j} 
    -\frac{1}{2}a^{3}\delta^{ij}\alpha_{j}\dot{\Phi}_{0}\dot{\Phi}_{0} 
    }}
    \,,
    \\\\
    &
    2(2c_{2}-c_{3})a^{2}\epsilon^{ijk}\partial_{j}{\dot{\alpha}_{k}} 
    -2(-2c_{2}+c_{3})a\delta^{ij}\Delta{\tilde{V}_{j}} 
    = 
    \\
    &\qquad
    \underset{\mathrm{FLRW\,contribution}}{\underbrace{
    -4(-2c_{2}+3c_{3})a^{2}H\epsilon^{ijk}\partial_{k}\alpha_{j} 
    +2(-2c_{2}+3c_{3})a^{2}H\epsilon^{ijk}\partial_{j}G_{k} 
    +6(2c_{2}-3c_{3})a^{3}H^{2}\delta^{ij}\tilde{V}_{j} 
    }}
    \\
    &\qquad
    \underset{\mathrm{matter\,contribution}}{\underbrace{
    +2a^{3}V_{0}\delta^{ij}\tilde{V}_{j} 
    -a^{3}\delta^{ij}\tilde{V}_{j}\dot{\Phi}_{0}\dot{\Phi}_{0} 
    -4(-2c_{2}+3c_{3})a^{3}\dot{H}\delta^{ij}\tilde{V}_{j} 
    }}
    \,.
\end{split}
\label{EoMs for Type 3}
\end{equation}
Notice that the second and third equations constitute six constraints in total, as implied by the kinetic matrix given in Eq.~\eqref{kinetic matrix in Type 3}. 
We consider two cases; 
Case 1. Vacuum Minkowski spacetime limit ($\Phi_{0}=0$, $V_{0}=0$, and $H=0$); 
Case 2. Vacuum FLRW spacetime limit ($\Phi_{0}=0$, $V_{0}=0$, and $H=\mathrm{Const.}$; see Eq.~\eqref{simplified bcgd eqs of NGR with test matter filed}).~\footnote{
One can investigate the Minkowski background case together with a scalar matter field, \textit{i.e.}, $\Phi_{0}\neq0$, $V_{0}\neq0$, $H=0$, in the same manner. 
However, we do not explicitly describe this analysis here since this case is irrelevant to our concern. 
The result shows that the matter-induced vector propagating modes emerge.  
} 

\subsection{Case 1. Vacuum Minkowski spacetime limit: $\Phi_{0}=0$, $V_{0}=0$, and $H=0$}

The equations of motion, Eq.~\eqref{EoMs for Type 3}, reduce to 
\begin{equation}
\begin{split}
    &
    -2(-2c_{2}+c_{3})a^{3}\delta^{ij}\ddot{\alpha}_{j} 
    -2c_{2}a\delta^{ij}\Delta\alpha_{j} 
    +2c_{2}a\delta^{ij}\Delta G_{j} 
    +2(2c_{2}-c_{3})a^{2}\epsilon^{jki}\partial_{k}\dot{\tilde{V}}_{j} 
    =
    0
    \,,
    \\\\
    &
    2c_{2}a\delta^{ij}\Delta{\alpha_{j}} 
    -2c_{2}a\delta^{ij}\Delta{G_{j}} 
    = 
    0
    \,,
    \\\\
    &
    2(2c_{2}-c_{3})a^{2}\epsilon^{ijk}\partial_{j}{\dot{\alpha}_{k}} 
    -2(-2c_{2}+c_{3})a\delta^{ij}\Delta{\tilde{V}_{j}} 
    = 
    0
    \,.
\end{split}
\label{Case 1 EoMs for Type 3}
\end{equation} 
These equations correspond, in the same order, to Eq.~(12) in Ref.~\cite{Golovnev:2026xjc} under the restriction $a=b$ and the identification of perturbative variables, $\alpha_{i}\leftrightarrow u_{i}$, $G_{i}\leftrightarrow v_{i}$, and $\tilde{V}_{i}\leftrightarrow w_{i}$, up to a difference in parametrization. 
The author of Ref.~\cite{Golovnev:2026xjc} then attempts to find for $\dot{w}_{i}$ from the second equation and substitute it into the third equation in order to recover Eq.~(7) in Ref.~\cite{Golovnev:2026xjc}. 
However, this procedure does not work, since the term containing $\dot{w}_{i}$ in the second equation vanishes when $a=b$ is imposed. 
In other words, this manipulation is too general to correctly handle the equations of motion in Type~3. 
Consequently, Eq.~(7) in Ref.~\cite{Golovnev:2026xjc} cannot be reproduced by this procedure in Type~3. 

We solve Eq.~\eqref{Case 1 EoMs for Type 3}. 
Substituting the second equation into the first one, we obtain 
\begin{equation}
    a\delta^{ij}\ddot{\alpha}_{j}  
    +
    \epsilon^{ijk}\partial_{k}\dot{\tilde{V}}_{j} 
    =
    0
    \,.
\label{hybrid of first EoM with second EoM in Case 1}
\end{equation}
On the other hand, the third equation of motion reduces to 
\begin{equation}
\begin{split}
    a\epsilon_{ijk}{\dot{\alpha}}{}^{k} 
    + 
    \partial_{j}{\tilde{V}_{i}}
    =
    D_{ij}
    \,,
\end{split}
\label{integrated third equation in Case 1}
\end{equation}
where $D_{ij}$ is an auxiliary spatial tensor satisfying $\partial_{i}D^{i}{}_{j}=0$ and $\dot{D}_{ij}=0$. 
Substitute this equation into Eq.~\eqref{hybrid of first EoM with second EoM in Case 1}, we obtain 
\begin{equation}
\begin{split}
    -a\ddot{\alpha}_{i}
    =
    0
    \,.
\end{split}
\end{equation}
This equation has a solution: $\alpha_{i}=a_{i}t+b_{i}$ for arbitrary spatial vectors $a_{i}$ and $b_{i}$. 
This result is consistent with the rank of the kinetic matrix given in Eq.~\eqref{kinetic matrix in Type 3}. 
However, this solution cannot be applicable to cosmology since it is not an oscillatory one; 
the absolute value of this solution becomes arbitrarily large when time passes. 
Therefore, $a_{i}$ should be zero, and this solution does not propagates. 

\subsection{Case 2. Vacuum FLRW background spacetime limit: $\Phi_{0}=0$, $V_{0}=0$, and $H=\mathrm{Const.}$}

In the case of vacuum but FLRW background spacetime, the situation changes. 
The equations of motion, Eq.~\eqref{EoMs for Type 3}, reduce to 
\begin{equation}
\begin{split}
    &
    -2(-2c_{2}+c_{3})a^{3}\delta^{ij}\ddot{\alpha}_{j} 
    -2c_{2}a\delta^{ij}\Delta\alpha_{j} 
    +2c_{2}a\delta^{ij}\Delta G_{j} 
    +2(2c_{2}-c_{3})a^{2}\epsilon^{jki}\partial_{k}\dot{\tilde{V}}_{j} 
    =
    \\
    &\qquad
    \underset{\mathrm{FLRW\,contribution}}{\underbrace{
    +6(-2c_{2}+c_{3})Ha^{3}\delta^{ij}\dot{\alpha}_{j} 
    +6(-3c_{2}+3 c_{3})a^{3}H^{2}\delta^{ij}\alpha_{j} 
    -2(-2c_{2}+3c_{3})a^{3}H\delta^{ij}\dot{G}_{j} 
    -6(-2c_{2}+c_{3})a^{3}H^{2}\delta^{ij}G_{j} 
    }}
    \,,
    \\\\
    &
    2c_{2}a\delta^{ij}\Delta{\alpha_{j}} 
    -2c_{2}a\delta^{ij}\Delta{G_{j}} 
    = 
    \\
    &\qquad
    \underset{\mathrm{FLRW\,contribution}}{\underbrace{
    +2(-2c_{2}+3c_{3})a^{3}H\delta^{ij}\dot{\alpha}_{j} 
    +12c_{3}a^{3}H^{2}\delta^{ij}\alpha_{j} 
    -24c_{2}a^{3}H^{2}\delta^{ij}G_{j} 
    -2(-2c_{2}+3c_{3})a^{2}H\epsilon^{ijk}\partial_{j}{\tilde{V}_{k}} 
    }}
    \,,
    \\\\
    &
    2(2c_{2}-c_{3})a^{2}\epsilon^{ijk}\partial_{j}{\dot{\alpha}_{k}} 
    -2(-2c_{2}+c_{3})a\delta^{ij}\Delta{\tilde{V}_{j}} 
    = 
    \\
    &\qquad
    \underset{\mathrm{FLRW\,contribution}}{\underbrace{
    -4(-2c_{2}+3c_{3})a^{2}H\epsilon^{ijk}\partial_{k}\alpha_{j}
    +2(-2c_{2}+3c_{3})a^{2}H\epsilon^{ijk}\partial_{j}G_{k} 
    +6(2c_{2}-3c_{3})a^{3}H^{2}\delta^{ij}\tilde{V}_{j}
    }}
    \,.
\end{split}
\label{Case 2 EoMs for Type 3}
\end{equation}
Substituting the second equation into the first one, we obtain 
\begin{equation}
    a\delta^{ij}\ddot{\alpha}_{j} 
    +
    \epsilon^{ijk}\partial_{k}\dot{\tilde{V}}_{j} 
    =
    A_{i}(\alpha_{i},\dot{\alpha}_{i},G_{i},\dot{G}_{i},\tilde{V}_{i};a,H) 
    \,.
\label{hybrid EoM of first with second EoM in Case 2}
\end{equation}
where we have set 
\begin{equation}
\begin{split}
    &A_{i}(\alpha_{i},\dot{\alpha}_{i},G_{i},\dot{G}_{i},\tilde{V}_{i};a,H) 
    \\
    &
    :=
    -\frac{H}{2a^{2}(-2c_{2}+c_{3})}\Big{[}
    4(-4c_{2}+3c_{3})a^{3}\delta^{ij}\dot{\alpha}_{j} 
    +6(-3c_{2}+5c_{3})a^{3}H\delta^{ij}\alpha_{j} 
    -2(-2c_{2}+3c_{3})a^{3}\delta^{ij}\dot{G}_{j} 
    \\
    &\qquad\qquad\qquad\qquad\qquad
    -6(2c_{2}+c_{3})a^{3}H\delta^{ij}G_{j} 
    -2(-2c_{2}+3c_{3})a^{2}\epsilon^{ijk}\partial_{j}{\tilde{V}_{k}} 
    \Big{]}
    \,.
\end{split}
\end{equation}
We remark that when the Minkowski limit, $H\rightarrow0$, Eq.~\eqref{hybrid EoM of first with second EoM in Case 2} reproduces Eq.~\eqref{hybrid of first EoM with second EoM in Case 1}. 
The third equation of motion reduces to 
\begin{equation}
    a\epsilon_{ijk}\dot{\alpha}^{k}
    +
    \partial_{j} \tilde{V}_{i} 
    = 
    \partial_{i}\Phi_{j} + D_{ij}
    \,,
\label{integrated third equation in Case 2}
\end{equation}
where $D_{ij}$ is an auxiliary spatial tensor satisfying $\partial_{i}D^{i}{}_{j}=0$, $\dot{D}_{ij}=0$, and $\Phi_{i}$ is a solution of the following Poisson equation:
\begin{equation}
\begin{split}
    &\Delta \Phi_{i}
    =
    B_{i}(\alpha_{i},G_{i},\tilde{V}_{i};a,H)
    \,,
    \\
    &
    B_{i}(\alpha_{i},G_{i},\tilde{V}_{i};a,H)
    \\
    &
    :=
    -\frac{H}{2a(-2c_2+c_3)}\Big{[}
    -4(-2c_{2}+3c_{3})a^{2}\epsilon^{ijk}\partial_{k}\alpha_{j} 
    +2(-2c_{2}+3c_{3})a^{2}\epsilon^{ijk}\partial_{j}G_{k} 
    +6(2c_{2}-3c_{3})a^{3}H\delta^{ij}\tilde{V}_{j} 
    \Big{]}
    \,,
\end{split}
\label{potentian of B}
\end{equation}
or a solution $\Phi_{i}$ of this equation is explicitly given by 
\begin{equation}
    \Phi_{i}(\vec{x},t)
    =
    -\frac{1}{4\pi}
    \int_{\mathbb{R}^3}
    d^3x'
    \,
    \frac{B_{i}(\alpha_{i},G_{i},\tilde{V}_{i};a,H)}{|\vec{x}-\vec{x}'|}
    +
    \Phi_{i}^\mathrm{(Minkowski\,limit)}(\vec{x},t)
    \,.
\label{solusion of Phi}
\end{equation}
When the Minkowski limit, $H\rightarrow0$, Eq.~\eqref{potentian of B} reduces to $\Delta\Phi_{i}=\Delta\Phi_{i}^\mathrm{(Minkowski\,limit)}=0$. 
This equation implies that $\Phi_{i}=\Phi_{i}^\mathrm{(Minkowski\,limit)}=0$ under the boundary condition: $\Phi_{i}^\mathrm{(Minkowski\,limit)}\rightarrow0 \quad(|\vec{x}|\rightarrow\infty)$. 
Thus, Eq.~\eqref{integrated third equation in Case 2} reproduces Eq.~\eqref{integrated third equation in Case 1}. 

Substituting Eq.~\eqref{integrated third equation in Case 2} into Eq.~\eqref{hybrid EoM of first with second EoM in Case 2}, we obtain 
\begin{equation}
    - a\ddot{\alpha}_{i} 
    - 
    A_{i}(\alpha_{i},\dot{\alpha}_{i},G_{i},\dot{G}_{i},\tilde{V}_{i};a,H) 
    + 2a^{3}H{\dot{\alpha}_{i}} 
    + a^{2}\epsilon_{ijk}\partial^{j}\dot{\Phi}^{k} 
    =
    0
    \,.
\label{}
\end{equation}
Since $B$ contains only no time derivative terms, $\dot{\Phi}_{i}$ consists of terms up to first order time derivative. 
Therefore, we can rewrite this equation schematically as follows:
\begin{equation}
    \ddot{\alpha}_{i} + \mathscr{A}(\alpha_{i},G_{i},\dot{G}_{i},\tilde{V}_{i};a,H;\partial_{i})\dot{\alpha}_{i} + \mathscr{A}_{i}(\alpha_{i},G_{i},\dot{G}_{i},\tilde{V}_{i};a,H) 
    = 
    0 
    \,.
\end{equation}
Although this equation is difficult to solve analytically, it is possible to admit an oscillatory solution under an adequate dispersion relation. 
Therefore, we conclude that Type~3 can generically contain the background-induced propagating vector modes, and this result is consistent with the rank of the kinetic matrix given in Eq.~\eqref{kinetic matrix in Type 3}. 

\bibliography{Bibliography}
\bibliographystyle{utphys}

\end{document}